\documentclass[11pt,prd,superscriptaddress,nofootinbib]{revtex4}
\usepackage[english]{babel}
\usepackage{graphicx}
\usepackage{mathtext}
\usepackage{indentfirst}
\usepackage{epsfig,amsmath,amsfonts}
\usepackage{color}

\def\xx{{\boldsymbol\xi}}

\newcommand{\beq}[1]{\begin{eqnarray}\label{#1}}
\newcommand\eeq {\end{eqnarray}}
\newcommand\bqa {\begin{eqnarray}}
\newcommand\eqa {\end{eqnarray}}
\newcommand\pr {\partial}

\newcommand{\bear}{\begin{array}}
\newcommand{\enar}{\end{array}}

\def\n{\kappa}
\def\Wl{W_{\lambda}}
\def\wl{w_{\lambda}}

\def\bC{{\bf C}}
\def\bR{{\bf R}}

\def\ch{\mathop{\rm ch}\nolimits}
\def\sh{\mathop{\rm sh}\nolimits}

\def\CC{{\cal C}}

\def\TT{{\cal T}}

\def\interior#1{\setbox1=\hbox{$#1$}\rlap{$#1$}\kern0.4\wd1\raise1.1\ht1%
\hbox{$\scriptstyle \circ$}}

\def\boxit#1#2{\setbox1=\hbox{\kern#1{#2}\kern#1}%
\dimen1=\ht1 \advance \dimen1 by #1 \dimen2=\dp1 \advance \dimen2 by #1
\setbox1=\hbox{\vrule height\dimen1 depth\dimen2\box1\vrule}%
\setbox1=\vbox{\hrule\box1\hrule}%
\advance \dimen1 by .4pt \ht1=\dimen1 \advance \dimen2 by .4pt \dp1=\dimen2
\box1\relax}
\def\endprf{\raise .5ex\hbox{\boxit{2pt}{\ }}}

\def\ifundefined#1{\expandafter\ifx\csname#1\endcsname\relax}

\def\beq{\begin{equation}}
\def\endq{\end{equation}}
\def\beqa{\begin{eqnarray}}
\def\endqa{\end{eqnarray}}
\def\x{{\bf x}}

\renewcommand{\cosh}{\ch}\renewcommand{\sinh}{\sh}

\begin{document}

\hfill ITEP-TH-5/17

\vspace{5mm}

\centerline{\Large \bf Infrared dynamics  of massive scalars from the complementary series}
\centerline{\Large \bf in de Sitter space}

\vspace{5mm}

\centerline{E. T. ${\rm Akhmedov}^{1, 2}$, U. ${\rm Moschella}^{3}$, K. E. ${\rm Pavlenko}^{1, 2}$ and F. K. ${\rm Popov}^{1, 2, 4}$}

\begin{center}
{\it $\phantom{1}^{1}$ B. Cheremushkinskaya, 25, Institute for Theoretical and Experimental Physics, 117218, Moscow, Russia}
\end{center}

\begin{center}
{\it $\phantom{1}^{2}$ Institutskii per, 9, Moscow Institute of Physics and Technology, 141700, Dolgoprudny, Russia}
\end{center}

\begin{center}
{\it $\phantom{1}^{3}$ Universit\`a degli Studi dell'Insubria - Dipartimento DiSAT, Via
Valleggio 11 - 22100 Como - Italy and
INFN, Sez di Milano, Via Celoria 16, 20146, Milano - Italy}
\end{center}

\begin{center}
{\it $\phantom{1}^{4}$ Joseph Henry Laboratories, Princeton University, Princeton, NJ 08544, USA}
\end{center}

\vspace{3mm}



\centerline{\bf Abstract}

We continue a previous study  about the infrared loop effects
in the $D$-dimensional de Sitter space for a
real scalar $\phi^4$ theory from the complementary series whose bare mass belongs to the interval $\frac{\sqrt{3}}{4}\, \left(D-1\right) < m \leq \frac{D-1}{2}$, in units of the Hubble scale. The lower bound comes from the appearance of discrete states in the mass spectrum of the theory when that bound is violated, causing large IR loop effects in the vertices. We derive an equation which allows to perform a self--consistent resummation of the leading IR contributions from all loops to the two-point correlation functions in an expanding Poincar\'{e} patch of the de Sitter manifold.  The resummation can be done for density perturbations of the Bunch--Davies state which violate the de Sitter isometry. There exist solutions having a singular (exploding) behavior and therefore the backreaction can change the de Sitter geometry.

\vspace{10mm}

\section{Introduction}

Quantum effects in de Sitter (dS) spacetime have received considerable attention in recent times \cite{Antoniadis:2006wq}---\cite{Akhmedov:2013vka}.  These  effects  are {rather} different from those in flat or anti-de Sitter space \cite{AkhmedovSadofyev}. As is shown in \cite{PolyakovKrotov}---\cite{AkhmedovPopovSlepukhin}, the peculiar infrared (IR) behavior of interacting nonconformal fields in dS space is that  there are large IR loop corrections or divergences  even for very massive fields  for any initial state  (see \cite{Akhmedov:2013vka} for a review). The quantum corrections {eventually become} of the same order or even dominate the tree-level contributions and the situation is similar to the one encountered in  non-stationary condensed matter theory \cite{LL,Kamenev} (see also \cite{Akhmedov:2014hfa,Akhmedov:2014doa} for the secular loop effects in strong electric fields in QED and \cite{Akhmedov:2015xwa} for the secular growth of loop corrections to the Hawking radiation).

In this note we consider a real, massive, minimally coupled scalar field theory:

\bqa\label{freeac}
S = \int d^Dx \sqrt{|g|}\, \left[\frac12 \, g^{\alpha\beta} \, \pr_\alpha \phi \, \pr_\beta \phi + \frac12 \, m^2 \, \phi^2 + \frac{\lambda}{4!} \, \phi^4\right].
\eqa
The theory in question is restricted to the expanding Poincar\'{e} patch (EPP) of dS space:

\bqa\label{PP}
ds^2 = \frac{1}{\eta^2}\,\left[- d\eta^2 + d\vec{x}^2\right], \quad \eta = e^{-t},
\eqa
where the conformal time is ranging form $\eta = + \infty$ at past infinity ($t=-\infty$) to $\eta = 0$ at future infinity ($t=+\infty$). Throughout this note we set the radius of the dS spacetime to one. 
Our goal is to check whether the assumption of negligible backreaction
is self-consistent or not for various sorts of initial conditions.

The EPP coordinate system has a well known peculiarity:  due to the presence of the conformal factor $1/\eta^2$  multiplying the spatial part of the metric every wave experiences an IR shift towards future infinity,  i.e. future infinity of the EPP  corresponds to the IR limit of the physical momentum, while past infinity corresponds to the UV limit. Correspondingly, a generic free scalar mode in the EPP has the following properties \cite{Chernikov:1968zm,Bunch:1978yq}: first, due to spatial flatness of the EPP a scalar mode may be factorized in terms of plane waves as follows:  $\phi_p(\eta, \vec{x}) = \eta^{(D-1)/2}\,h(p\eta)\, e^{-i\,\vec{p}\, \vec{x}}$, where $h(p\eta)$ is a solution of the Bessel equation of order $\nu = \sqrt{\left(\frac{D-1}{2}\right)^2-m^2}$. Second, any Bessel function of this order behaves as follows:

\bqa\label{beha}
h(p\eta) = \left\{\begin{array} {c} A \, \frac{e^{i\, p\eta}}{\sqrt{p\eta}} + B \, \frac{e^{- i\, p\eta}}{\sqrt{p\eta}}, \quad p\eta \gg \left|\nu\right| \\
C \, (p\eta)^{\nu} + D \, (p\eta)^{-\nu}, \quad p\eta \ll \left|\nu\right|. \end{array} \right. \label{aa}
\eqa
Here $A,B,C,D$ are complex constants which are fixed by the canonical commutation relations and some other additional criterion. Due to the symmetries of the EPP one cannot disentangle the comoving momentum $p$ and conformal time $\eta$; all physical quantities depend on the combination $p\eta$ which is referred to as the {\em physical momentum}.
Near past infinity of the chosen EPP the  physical momentum  $p\eta$ tends to infinity and every mode behaves asymptotically as a plain wave in flat spacetime.  This is because high energy modes are not sensitive to the comparatively small curvature of the background. One can actually introduce  in that region a notion of particle with positive energy, because the free Hamiltonian can be diagonalized there.  Thus, at past infinity of the EPP the background gravitational field is effectively switched off.
Equivalently, for a given mode function $\phi_p$ behaves as a plain wave in flat space
when $p\eta \gg \left|\nu\right|$.  On the other hand for low physical momenta the behavior of the modes (see Eq. \ref{beha}) is very much different from the one in flat space. It is exactly the latter region of physical momenta $p \eta \ll \left|\nu\right|$ that generates large IR corrections to correlation functions.

This is, roughly speaking, the origin of the strong IR effects in dS space that are discussed in the present paper.

As one can notice from (\ref{beha}) the character of the behavior of the mode functions (and, hence, of the IR effects \cite{AkhmedovPopovSlepukhin}) depends on whether $m$ is greater or smaller than $(D-1)/2$ in units of the dS radius. Scalar fields with $m > (D-1)/2$ are associated with the  principal series of representations of the dS group. In this case $\nu$ is purely imaginary and the mode functions (\ref{beha}) oscillate at small physical momenta. The IR physics in this case has been extensively studied in \cite{PolyakovKrotov}---\cite{Akhmedov:2013vka}.

Here we want to explore the case of light scalars (the complementary series).
Masses of such fields obey $0 < m \leq (D-1)/2$ in units of the dS radius.
Now $\nu$ is real and the mode functions (\ref{beha}) {\it do not} oscillate at future infinity of the EPP.

\subsection{The Schwinger--Keldysh formalism}

The action (\ref{freeac}) defines a field theory in a  non--stationary background (\ref{PP}).
Therefore, a perturbative expansion of the correlation functions should be constructed in terms of  three propagators \cite{LL} (see also \cite{Akhmedov:2013vka}, \cite{Kamenev}, \cite{vanderMeulen:2007ah}). Two of them are the standard retarded and advanced propagators; they are purely algebraic, i.e. they do not depend on the chosen Fock space realisation of the (free or tree level) theory
\bqa
D^{\frac{R}{A}}_0\left(\eta_1, \vec{x}_1; \eta_2, \vec{x}_2\right) = \pm \theta\left(\mp \Delta \eta_{12}\right) \, \left\langle \left[\phi\left(\eta_1, \vec{x}_1\right),\, \phi\left(\eta_2, \vec{x}_2\right)\right]\right\rangle, \quad \Delta \eta_{12} = \eta_1 - \eta_2,
\eqa
where $[\cdot, \cdot]$ is the commutator. The  Keldysh propagator is the "vacuum" expectation value of the anticommutator:
\bqa
D^K_0\left(\eta_1, \vec{x}_1; \eta_2, \vec{x}_2\right) = \frac{1}{2} \, \left\langle \left\{\phi\left(\eta_1, \vec{x}_1\right), \,\phi\left(\eta_2, \vec{x}_2\right)\right\}\right\rangle.
\eqa
The Keldysh propagator does depend on the Fock space realisation of the theory.

The EPP is invariant under space translations $\vec x \to \vec x + \vec a$.  We will only consider quantisations where space translations are unbroken. This means in particular that we assume that all the propagators depend on the difference vectors $\vec {x}_2 - \vec {x}_1$. It is therefore advantageous  to Fourier transform of all the quantities w.r.t. the above difference vectors\footnote{Note that due to the expansion of the EPP every spatially inhomogeneous perturbation fades away at future infinity. Thus, the IR effects under study, appearing from the future infinity region of the EPP, are not very much sensitive to such inhomogeneous perturbations. Hence, our methods are also applicable in the presence of such perturbations.}:
\begin{equation}
D^{K,R,A}_{0}\left(p\,|\eta_1, \eta_2\right) \equiv \int d^{D-1}x \, e^{-i\, \vec{p}\, \vec{x}} D^{K,R,A}_0(\eta_1, \vec{x}; \eta_2, 0).
\end{equation}
A partial Fourier transformation is also helpful to keep track separately of the behavior of each mode with a given physical momentum.
Here are the Fourier transformed tree-level retarded and advanced propagators  \cite{vanderMeulen:2007ah}:
\bqa\label{DRA}
D_{0}^{\frac{R}{A}}\left(p\,|\eta_1,\eta_2\right) = \pm \theta\left(\mp\Delta\eta_{12}\right) \,2\,(\eta_1\eta_2)^{\frac{D-1}{2}}\,{\rm Im}\left[h(p\eta_1)h^*(p\eta_2)\right].
\eqa
If the initial state $\left|\Psi \right\rangle$ respects the spatial translational invariance the (tree-level) Keldysh propagator can be written as follows:
\begin{gather}
D_0^K\left(p|\eta_1, \eta_2\right) = \nonumber \\ = \left(\eta_1 \eta_2\right)^{\frac{D-1}{2}} \, \left[\left(\frac12 + \left\langle \Psi, \, a^+_{\vec{p}} \, a_{\vec{p}}\, \Psi \right\rangle\right) \, h(p\eta_1) h^*(p\eta_2) + \left\langle \Psi, \,a_{\vec{p}} \, a_{-\vec{p}}\,\Psi \right\rangle \, h(p\eta_1) h(p\eta_2) + h.c.\right]. \label{eq2}
\end{gather}
In a ground state $a_{\vec{p}} \, \left|\Psi \right\rangle = 0$ the latter expression reduces to
\bqa\label{DK}
D_{0}^{K}\left(p\,|\eta_1,\eta_2\right) = (\eta_1\eta_2)^{\frac{D-1}{2}}\,{\rm Re}\left[h(p\eta_1)h^*(p\eta_2)\right].
\eqa
When the mass is non--zero
 there is a one-parameter family of dS invariant quantisations known as the $\alpha$-vacua
 (see \cite{Spindel,Allen:1985ux, Mottola:1984ar}). In all these cases the Keldysh propagator  $D_0^{K}(1,2)$ depends only on the invariant geodesic distance between the two points (while this happens for tree--level $D_0^{R,A}$ for any given state  $\left|\Psi \right\rangle$).

\section{On different types of secular effects. Results of this paper}

There are different sorts of secularly growing contributions in non-stationary situations in general, and in dS space in particular. To begin, there is a secular growth which is specific to dS space and is already present at tree--level (see e.g. \cite{Allen:1987tz,Tsamis:2005hd,bemtach,bemtach2}). It exists for all minimally coupled scalar tachyons, a family of fields whose squared mass is negative or zero --- it includes the massless scalar field. In these  cases canonical quantization gives rise to  two--point  functions that do not depend only on  the geodesic distance and there exists no Fock space representation where the representation of dS group is unitary (see \cite{bemtach,bemtach2} --- for masses equal to
$m^2 = -n(n+d-1)$ with $n$ a nonnegative integer, there exists however a construction similar to the Gupta-Bleuler quantization of the free photon field).

By subtracting the UV divergence one finds that the correlation functions grow with time; for instance for the massless scalar field one has  $D_0(t,\vec{x}) \propto \left|\log \eta\right| \sim t$ for the coincident points. Furthermore, taking into account loops  e.g. for a $\lambda \, \phi^4$ self-interaction,  the  secular growth is  $\Delta_n D(t, \vec{x}) \propto t \, \left(\lambda t^2\right)^n$ where $n$ grows with the number of loops.

Such a secular growth results in a breakdown of perturbation theory. In fact, for every  $\lambda$ however small, after a time $t$ long enough $\lambda t^2 \sim 1$ 
and quantum corrections become of the same order of the tree--level amplitude. A resummation of (at least) the leading  contributions from  loops is therefore mandatory for a meaningful perturbation calculation in dS space--time.

One possible scheme  to perform  the resummation has been proposed in \cite{Starobinsky}.
It makes use of the stochastic approach to quantum field theory  and allows in a certain limit the resummation of the secular corrections to the Bunch--Davies (BD) vacuum (or its analog for the massless scalar) in the EPP \cite{Tsamis:2005hd}.  After the resummation the dS invariance of the correlation functions in the massless scalar field theory is restored. 
The approach of \cite{Starobinsky} allows to control various sorts of secular contributions
(without disentangling them) for scalar fields with any nonnegative mass {belonging to the complementary series}. But it does this only in very special situations and the determination of the exact limits of validity of this approach is an important, separate, issue. In particular, for obvious reasons, this method cannot be applied to global dS space or to the contracting Poincar\'{e} patch. Also it is not applicable in the EPP for strong enough initial density perturbations of the BD state --- when there are strong non--linear correlations, even if they are spatially homogenous. The point is that the approach described in  \cite{Starobinsky} exploits a stochastic differential equation with a {\em linear} random source in a {\em non--linear} (self--interacting) theory. It is not applicable for the case when one has strong enough non--linearities. We will come back to this point below.

Another well known example of a secular IR loop correction in dS space is the following:   \begin{equation}
\Delta_n D\left(p|\eta_1, \eta_2\right) \propto \left(\eta_1\, \eta_2\right)^{(D-1)/2}\, \left[\lambda^2 \, \log \left(\eta_1/\eta_2\right)\right]^n = e^{-\frac{D-1}{2} \, \left(t_2+ t_1\right)} \, \lambda^{2 n \phantom{\frac12}} |t_1-t_2|^n
\end{equation} in the scalar $\lambda \, \phi^4$ theory,  when $\left|t_2 - t_1\right| \to \infty$.

Such a secular growth is quite universal and is present even for positive mass  \cite{Boyanovsky:1993xf,deVega:1997yw,Boyanovsky:1997cr}, \cite{Leblond}. Usually such a growth is caused by some imaginary contributions to the self--energy;
it is also present in Minkowski space--time, e.g. if one chooses an initial density matrix other than Planckian  and describes the instability of quasiparticles (in the latter case, however, the dS volume factor $e^{- \frac{D-1}{2} {\left(t_2+ t_1\right)}}$ is not present). This secular effect is present in all the propagators including the  retarded and advanced ones, and can be seen in their partially Fourier transforms. It may also be present in the vertices.

Some further comments are in order here  for understanding the physical origin and the differences between various types of secular effects. In flat space--time, in the standard non--stationary situation,  the Fourier transform of a propagator $D_0\left(p|t_1, t_2\right)$ is proportional to $e^{i\, \omega(p) \, \left(t_2 - t_1\right)}$ where $\omega(p)$ is the dispersion relation of the model under consideration.  The secular corrections to the self-energy are absorbed by the coefficient in front of the exponential  at every loop order. They  also depend on $t_2 - t_1$. After the resummation the growth in question at the leading order can be eliminated by a shift of the dispersion relation $\omega(p) \to \omega(p) + i \Gamma(p)$ or by a mass renormalization, when the contribution to the selfenergy is real. This growth cannot be attributed just to the IR effects in a proper sense, because it also appears in the UV domain.

Yet another secular effect at loop level is of the form
\begin{equation}
\Delta_n D^K\left(p|\eta_1, \eta_2\right) \propto \left(\eta_1\, \eta_2\right)^{(D-1)/2}\, \left[\lambda^2 \, \log \left(\frac{1}{\eta_1\eta_2}\right)\right]^n = e^{-\frac{D-1}{2} \, \left(t_2+ t_1\right)} \,\left[\lambda^2 \, (t_1+t_2)\right]^n
\end{equation}
which at leading order in $\lambda$ is present {\it only} in the Keldysh propagator when $t_2 + t_1 \to \infty$ and $t_2 - t_1 = const$ (in the conformal coordinates of the EPP in dS space the above condition is as follows: $p\, \sqrt{\eta_1 \eta_2} \to 0$,  with $\eta_1/\eta_2 = const$.) In this limit both points of the propagator are sent to  future infinity of the EPP while their coordinate time distance is held fixed. Such a growth is also universal to practically any non--stationary situation, including flat space non--stationary initial density matrix, strong electric fields in QED \cite{Akhmedov:2014hfa,Akhmedov:2014doa} and the Hawking radiation \cite{Akhmedov:2015xwa}.
Moreover, it is also present for massless scalars in dS space but is mixed with the other  contributions described above.

The origin of the $\lambda^2 (t_1+t_2)$ growth can be understood from the Keldysh propagator shown in Eq. (\ref{eq2}). In the interaction picture, which we use here, $a_{\vec{p}}$ and $a^+_{\vec{p}}$ are time independent in the Gaussian approximation. In this approximation all the time dependence is in the modes $\phi_p$. It means that in the Gaussian theory $\left\langle \Psi , a^+_{\vec{p}} \, a_{\vec{p}}\, \Psi \right\rangle$ and $\left\langle \Psi \,,a_{\vec{p}} \, a_{-\vec{p}}\,\Psi \right\rangle$ remain constant. In particular, if they have been chosen to be zero, they always remain zero.

But if one turns on selfinteractions  the above quantities, namely the population numbers and the anomalous quantum averages for the {\em exact modes}, start to depend on $t_1$ and $t_2$ and they receive the leading $\lambda^2 \, \left(t_2 + t_1\right)$ corrections in the limit under consideration\footnote{In general non--stationary situations correlation functions depend on each of their arguments separately, $D(t_1, t_2) = \overline{D}\left(t_1+ t_2, t_1 - t_2\right)$, rather than the distance between them, $D(t_2 - t_1)$: e.g. there are contributions to the propagator of the form $\Delta D(t_1, t_2) \propto n_p\left(\frac{t_1 + t_2}{2}\right) \, e^{- i \, \omega(p) \, \left(t_1 - t_2\right)}$ for some $n_p(t)$.}. This is  generally true for any non--stationary situation. 
Because of the different functional dependence this secular growth cannot be reabsorbed in a change of the dispersion relation  $\omega(p) \to \omega(p) + i \Gamma(p)$ or by a mass renormalization: $\lambda^2 \, \left(t_2 + t_1\right) \, e^{i\, \omega(p) \, \left(t_2 - t_1\right)} \neq e^{i\, \left(\omega(p) + \delta \omega\right)\, \left(t_2 - t_1\right)}$.
In this paper we focus on secular effects of the latter type. These effects can strongly affect the commonly accepted picture of the strength of stress--energy fluxes. In fact, the common wisdom is that quantum loop corrections lead only to UV renormalization while IR secular effects can just lead to  mass renormalization or to a mode decay rate.
Hence, according to the common wisdom  quantum corrections do not contribute to the stress--energy tensor\footnote{Note that in the strong background fields we calculate correlation functions and avoid using the notion of particles, unless it is appropriate for the interpretation of the energy--momentum flux \cite{Akhmedov:2013vka}. The latter one can be found from the correlation functions. Namely one can encounter situations in which there is a non--trivial energy flux, but there is no any suitable separation of it into particles -- into something that obeys the energy composition principle.} and one can safely apply the Gaussian approximation. In this approximation one sees only the amplification of the zero point fluctuations.

Indeed,  if loop corrections are irrelevant, one can use such a tree--level correlation functions as (\ref{DK}) to calculate the stress--energy tensor or the electric current: namely one can find in this way the Schwinger's current in strong electric fields, the Hawking energy flux in the gravitational collapse and Bunch--Davies expectation value of the stress--energy tensor in dS space. However, the time dependence of the level populations and the anomalous quantum averages for the {\em exact modes}, which we have just described, may drastically change the energy--momentum tensor (see \cite{Akhmedov:2014hfa,Akhmedov:2014doa,Akhmedov:2015xwa,Akhmedov:2013vka} for a number of examples). In fact, in this case to calculate the stress--energy tensor one has to use an analog of (\ref{eq2}), with time dependent $\langle a^+_p a_p\rangle$ and $\langle a_p a_{-p}\rangle$ which are attributed to the comoving volume.

\subsection{Complementary series}

For masses $m<(D-1)/2$  (the complementary series) generic modes behave as real powers of the conformal time (as opposite to oscillating imaginary powers, see Eq. (\ref{aa})).  As a result the dominant term in  $h(p\eta_1) \, h^*(p\eta_2)$ is proportional to  $\eta_2 \eta_1 = e^{-|\nu|(t_1 + t_2)}$. Hence, for light scalars the aforementioned universal secular effects, which are present even for massive fields, can be mixed up in the linear approximation to the Dyson--Schwinger equation. Moreover, because the modes do not oscillate, the imaginary contributions to the selfenergy may lead to a mass renormalization rather than to a decay rate in the same approximation. This fact makes it difficult to disentangle the different secular effects from each other, while it is necessary to disentangle them to understand quantum field theory in dS space, as we have explained in the previous paragraph. But this distinction is out of reach of the methods of \cite{Starobinsky}, \cite{Tsamis:2005hd}.

The situation regarding the fields of the complementary series has been explored in e.g. \cite{Gautier:2015pca}--\cite{Serreau:2013psa}.  The resummation of the secularly growing loops in the BD state has been attempted there by using different methods, including the solution of the Dyson--Schwinger equation in the large $N$ limit and the exact renormalization group equation for the IR cutoff. Their results agree with \cite{Starobinsky} and extend them to the case of non--coincident points of the two--point correlation functions. An interesting solution of the Dyson-Schwinger equations for the retarded, advanced and Keldysh propagators has been found. The approach adopted in \cite{Gautier:2015pca}, \cite{Gautier:2013aoa}, \cite{Serreau:2013psa} allows to resum simultaneously all the aforementioned different sorts of IR effects in the large $N$ limit without disentangling them.

However, in our opinion there remain certain unsatisfactory features which make the study of the dS light scalar field dynamics not yet complete.
The point is that the resummation is done  in  \cite{Gautier:2015pca}--\cite{Serreau:2013psa} only for the BD state and for the cases when the dS isometry is respected at every step of the quantization. It is quite understandable that in this case at the leading order quantum effects just lead to the renormalization of the cosmological constant and the mass of the field. But the question remains: what happens if one considers density perturbations of the BD state which (necessarily) violate the dS isometry? After all it is natural  to consider perturbations of a highly symmetric state and  trace their destiny: one should check whether they grow or fade away as the time goes by. That is what we propose to do in the present paper.

For massive scalars with exact  BD initial state at the light--like boundary of the EPP one can respect the dS isometry at every loop order \cite{Polyakov:2012uc} (see e.g. also \cite{Akhmedov:2013vka}). Moreover,  as we explain in the section VII, in this case the system of Dyson-Schwinger equations, which allows to resum leading loop corrections, reduces to a single linear integro--differential equation. For the complementary series similar equation was obtained in \cite{Gautier:2015pca}, \cite{Gautier:2013aoa}, \cite{Serreau:2013psa}. The difference of the situation in the latter references with respect to our case is that their equation was obtained in the large $N$ limit and resums a bit different sort of leading diagrams and IR effects, some of which are subleading in our approximation.

The situation with slight violations of the dS isometry is quite different. One cannot just put an initial comoving number density $n^0_p$ {\em for the exact modes} at past infinity (i.e. the light--like boundary) of the EPP, because then the physical density would become infinite. Due to the symmetries of the EPP the appropriate way of approaching the problem is as follows. As we have explained above, every physical quantity in the EPP is a function of the physical momentum, $P = p\eta$. Hence, one has to put an initial Cauchy condition for a given physical momentum, i.e. $n^0(P) = n\left(P_0\right)$, rather than just for initial conformal time $\eta_0$. A somewhat similar approach was adopted in  \cite{Starobinsky} and \cite{Tsamis:2005hd}. The initial density perturbation violates the dS isometry. The main difference with respect to the standard approach \cite{Starobinsky} is that here in general we obtain a {\em non--linear} integro--differential equation. For a very small initial density perturbation, when the initial comoving number density $n\left(P_0\right)$ is much smaller than 1, this equation can be reduced to a linear one, which is similar to the equation considered in \cite{Gautier:2015pca}, \cite{Gautier:2013aoa}, \cite{Serreau:2013psa}. However, if the initial perturbation is of order one, the full non--linear equation has explosive solutions.

Finally, let us remark that in dS space there can be secular IR effects in the vertices. They are very important for the resummation. The point is that to do the resummation one has to solve the system of Dyson--Schwinger equations for all propagators and vertices. As we explain in the main body of the text for lower masses higher and higher point functions become relevant in the IR limit\footnote{As we explain below, these secular effects in the vertices are present only for the fields whose mass is lower than certain bound, $m < \frac{\sqrt{3}}{4} \, (D-1)$. And as one lowers the mass higher and higher point correlation functions start to grow in the IR limit.}. This makes the problem under consideration practically impossible to solve, unless one can drop off the equations for the vertices and invent some suitable ansatz only for the two--point functions. That is possible to do only if the leading secular growth is present only in propagators and is absent in the vertices.

In our paper, we check the presence of the secular effects for the vertices and explain their physical origin. They appear due to the presence of bound states in the spectrum of the theory and signal that higher point correlations become relevant. We find the situation when these bound states and related secular effects in the vertices are absent. Then we perform the self-consistent resummation of the leading effects from all loops only when there are no secular effects in the vertices. We do that for an arbitrary, not necessary isometry respecting, density perturbations of the BD state. We derive an equation which provides the resummation, find its solutions and discuss their physical meaning.

\section{General discussion of the loop corrections to the propagators}

Let us start by discussing the  loop corrections to the Keldysh propagator $D^K(p|\eta_1,\eta_2)$ having chosen {an} initial  dS invariant state at past infinity of the EPP. It is not difficult to show that the massive $\lambda \, \phi^4$ theory does not possess any secularly growing contributions (of the last sort, which is described in the Introduction) to any propagator at the first-loop ``bubble'' diagram order ($\sim \lambda$). However, at the second-loop order ($\sim \lambda^2$) there is a large IR contribution to $D^K$ which is of interest for us. Two-loop diagrams that contain large IR corrections are of the "sunset" type:

\bqa\label{DSDK}
\Delta_2 D^{K}\left(p\, |\eta_1,\eta_2\right) = \frac{\lambda^2}{6} \int \frac{d^{D-1}\vec{q_1}}{(2\pi)^{D-1}}\frac{d^{D-1}\vec{q_2}}{(2\pi)^{D-1}} \iint_{+\infty}^0 \frac{d\eta_3 d\eta_4}{(\eta_3\eta_4)^D} \times \nonumber \\ \times \Biggl[ \Biggr.
3\,D_{0}^K\left(p\,|\eta_1,\eta_3\right) \, D_{0}^K\left(q_{1}|\eta_3,\eta_4\right) \, D_0^K\left(q_{2}\, |\eta_3,\eta_4\right) \,D_0^A\left(\left|\vec{p}-\vec{q}_1-\vec{q}_2\right|\, |\eta_3,\eta_4\right) \, D_0^A\left(p\, |\eta_4,\eta_2\right)
- \nonumber \\-
\frac{1}{4}\,D_{0}^K\left(p\,|\eta_1,\eta_3\right) \, D_{0}^A\left(q_{1}|\eta_3,\eta_4\right) \, D_0^A\left(q_{2}\, |\eta_3,\eta_4\right) \,D_0^A\left(\left|\vec{p}-\vec{q}_1-\vec{q}_2\right|\, |\eta_3,\eta_4\right) \, D_0^A\left(p\, |\eta_4,\eta_2\right)
- \nonumber \\-
\frac{3}{4}\,D_{0}^R\left(p\,|\eta_1,\eta_3\right) \, D_{0}^K\left(q_{1}|\eta_3,\eta_4\right) \, D_0^A\left(q_{2}\, |\eta_3,\eta_4\right) \,D_0^A\left(\left|\vec{p}-\vec{q}_1-\vec{q}_2\right|\, |\eta_3,\eta_4\right) \, D_0^A\left(p\, |\eta_4,\eta_2\right)
+ \nonumber \\+
D_{0}^R\left(p\,|\eta_1,\eta_3\right) \, D_{0}^K\left(q_{1}|\eta_3,\eta_4\right) \, D_0^K\left(q_{2}\, |\eta_3,\eta_4\right) \,D_0^K\left(\left|\vec{p}-\vec{q}_1-\vec{q}_2\right|\, |\eta_3,\eta_4\right) \, D_0^A\left(p\, |\eta_4,\eta_2\right)
- \nonumber \\-
\frac{3}{4}\,D_{0}^R\left(p\,|\eta_1,\eta_3\right) \, D_{0}^K\left(q_{1}|\eta_3,\eta_4\right) \, D_0^R\left(q_{2}\, |\eta_3,\eta_4\right) \,D_0^R\left(\left|\vec{p}-\vec{q}_1-\vec{q}_2\right|\, |\eta_3,\eta_4\right) \, D_0^A\left(p\, |\eta_4,\eta_2\right)
- \nonumber \\-
\frac{1}{4}\,D_{0}^R\left(p\,|\eta_1,\eta_3\right) \, D_{0}^R\left(q_{1}|\eta_3,\eta_4\right) \, D_0^R\left(q_{2}\, |\eta_3,\eta_4\right) \,D_0^R\left(\left|\vec{p}-\vec{q}_1-\vec{q}_2\right|\, |\eta_3,\eta_4\right) \, D_0^K\left(p\, |\eta_4,\eta_2\right)
+ \nonumber \\+
3\,D_{0}^R\left(p\,|\eta_1,\eta_3\right) \, D_{0}^R\left(q_{1}|\eta_3,\eta_4\right) \, D_0^K\left(q_{2}\, |\eta_3,\eta_4\right) \,D_0^K\left(\left|\vec{p}-\vec{q}_1-\vec{q}_2\right|\, |\eta_3,\eta_4\right) \, D_0^K\left(p\, |\eta_4,\eta_2\right)
\Biggl. \Biggr].
\eqa
Below we do not consider UV divergences but assume some kind of UV renormalization, i.e. that masses of the fields and coupling constants have been set equal to their physical renormalized values.
It is probably worth stressing here that mixed expressions where the partial Fourier transformation has been taken w.r.t. the spatial coordinates, are not sensitive to the UV divergences. In fact, to reveal the latter one needs an extra integration in the vertices: namely, it is necessary to transform back to the spacetime variables $(\eta, \vec{x})$. The leading IR contributions to $\Delta_2 D^K\left(p\, |\eta_1,\eta_2\right)$ are hidden within the following expression \cite{Akhmedov:2013vka}:

\bqa\label{EPP}
\Delta_2 D^{K}\left(p\, |\eta_1,\eta_2\right) \approx (\eta_1\,\eta_2)^{\frac{D-1}{2}}\, \left[h(p\eta_1)\, h^*(p\eta_2) \, n_2(p\eta) + h(p\eta_1)\, h(p\eta_2) \,\kappa_2(p\eta)^{\phantom{\frac12}} + c.c. \right],
\eqa
where $\eta   = e^{-t} = \sqrt{\eta_1\, \eta_2} = e^{-\frac{t_1+t_2}2}$ is the average conformal time and

\bqa
&& n_2(p\eta) = \frac{\lambda^2}{3} \int \limits^\eta_\infty \int \limits^\eta_\infty d\eta_3 d\eta_4 \, h(p\eta_3) h^*(p\eta_4) F(p,\eta_3,\eta_4), \nonumber \\
&& \kappa_2(p\eta) =- \frac{2\lambda^2}{3} \int \limits^\eta_\infty \int \limits^{\eta_3}_\infty d\eta_3 d\eta_4 h^*(p\eta_3) h^*(p\eta_4) F(p,\eta_3,\eta_4), \nonumber \\
&& F(p,\eta_3,\eta_4) =  \int \frac{d^{D-1} q_1}{(2\pi)^{D-1}}\frac{d^{D-1} q_2}{(2\pi)^{D-1}}(\eta_3 \eta_4)^{D-2}\times\nonumber\\
&& \times h(q_1\eta_4)h^*(q_1 \eta_3)h(q_2\eta_4)h^*(q_2 \eta_3)h\left(\left|\vec{p} - \vec{q}_1-\vec{q}_2\right|\eta_4\right)h^*\left(\left|\vec{p} - \vec{q}_1-\vec{q}_2\right|\eta_3\right),
\eqa
where subscript $2$ in $n_2$ and  $\kappa_2$ denotes the second loop contribution.

In deriving the above representation for $\Delta_2 D^K$ from  (\ref{DRA}), (\ref{DK}) and (\ref{DSDK})  in the limit $p\, \sqrt{\eta_1 \eta_2} \to 0$ with $ \eta_1/\eta_2 = const$, we neglected  the difference between $\eta_1$ and $\eta_2$ and replaced both of them by the average conformal time $\eta$ in the arguments of the Heaviside $\theta$-functions inside $D^{R,A}$ (see \cite{Akhmedov:2013vka} for more details).

In the following  we will estimate (\ref{EPP}) and see that for generic modes $h$ these quantities grow as $p\eta \to 0$ even when zero values of $n_0$ and $\kappa_0$ had been chosen at past infinity (see the tree--level expressions for $D^K$).

The character of the two-loop corrections to the retarded and advanced propagators depends neither on the choice of the mode functions $h$ nor on the mass of the field. It is the same as in the case of the principal series \cite{Leblond,Akhmedov:2013vka} (see also \cite{Kamenev} for a more general discussion). In fact, the two-loop contribution to $D^R$ is as follows:

\bqa\label{oneloopR}
\Delta_2 D^R\left(p\,|\eta_1, \eta_2\right) = \frac{\lambda^2}{6} \, \int \frac{d^{D-1}\vec{q}_1}{\left(2\, \pi\right)^{D-1}} \, \int \frac{d^{D-1}\vec{q}_2}{\left(2\, \pi\right)^{D-1}} \, \iint_{+\infty}^0 \frac{d\eta_3\, d\eta_4}{\left(\eta_3\, \eta_4\right)^{D}} \times \nonumber \\ \times
\Biggl[ \Biggr.
3\,D_{0}^R\left(p\,|\eta_1,\eta_3\right) \, D_{0}^R\left(q_{1}|\eta_3,\eta_4\right) \, D_0^K\left(q_{2}\, |\eta_3,\eta_4\right) \,D_0^K\left(\left|\vec{p}-\vec{q}_1-\vec{q}_2\right|\, |\eta_3,\eta_4\right) \, D_0^R\left(p\, |\eta_4,\eta_2\right)
- \nonumber \\-
\frac{1}{4}\,D_{0}^R\left(p\,|\eta_1,\eta_3\right) \, D_{0}^R\left(q_{1}|\eta_3,\eta_4\right) \, D_0^R\left(q_{2}\, |\eta_3,\eta_4\right) \,D_0^R\left(\left|\vec{p}-\vec{q}_1-\vec{q}_2\right|\, |\eta_3,\eta_4\right) \, D_0^R\left(p\, |\eta_4,\eta_2\right)\ \Biggl.\Biggr].
\eqa
Due to the presence of $D^R$ inside the loop and in the external legs, the limits of integration over $\eta_{3,4}$ are such that $\eta_1 > \eta_3 > \eta_4 > \eta_2$. As the result, the integral (\ref{oneloopR}) does not contain growing corrections when $\eta_1/\eta_2$ is held fixed. The situation with the advanced propagator is the same.

We will discuss loop corrections to the vertices below for specific choices of $h$ separately.

\section{Bunch-Davies fields.}

In this section we examine the BD fields of the complementary series.
The BD modes are proportional to the Hankel functions: $h(x) \propto H^{(1)}_\nu(x) = J_\nu(x) + i \, Y_\nu(x)$; $h^*$ is just the complex conjugate of $h$ \cite{Bunch:1978yq}. When $p\eta \to \infty$ they behave as $h(p\eta) \sim \frac{e^{i\, p\eta}}{\sqrt{p\eta}}$ and represent pure waves at past infinity of the EPP or in--modes. These modes diagonalize the free Hamiltonian at past infinity. On the other hand, when $p\eta \to 0$ they behave as $h(p\eta) \approx A_- \, (p\eta)^{-\nu} + i \, A_+\, (p\eta)^{\nu} + B \, (p\eta)^{-\nu + 2}$ where $A_\pm$ and $B$ are  real constants. In this paper we consider fields from the complementary series, i.e. $\nu$ is real and $0 < \nu < (D-1)/2$. We keep the $B\, x^{-\nu + 2}$ term because this term will dominate over $A_+\, x^\nu$, as $x\to 0$, if $\nu > 1$ and the presence of $A_+$ is important as we will see below. All that we say in our paper is valid for $\nu \leq 1$, but our discussion can be straightforwardly extended to the case of $1< \nu < (D-1)/2$ (for $D > 3$), if we take into account the $B$ term in $h$. 

\subsection{Corrections to the Keldysh propagator}
\label{IIIA}
Unlike the case of the principal series \cite{Akhmedov:2013vka}, the function $F(p,\eta_3,\eta_4)$ in (\ref{EPP}) may contain large contributions as $p\to 0$. In order to estimate this function let us  divide the area of integration over $d^{D-1}q_{1,2}$ into four regions: a region  where $\left|\vec{q}_{1,2}\right| \lesssim \left|\vec{p}\right|$, another  where $\left|\vec{q}_{1,2}\right| \gtrsim \left|\vec{p}\right|$ and two other  regions where $\left|\vec{q}_1\right| \lesssim \left|\vec{p}\right| \lesssim \left|\vec{q}_2\right|$ or  $\left|\vec{q}_2\right| \lesssim \left|\vec{p}\right| \lesssim \left|\vec{q}_1\right|$.  Here we present only vague estimates, in the next section we discuss the origin of these complications more rigorously and explain their physical meaning.

To estimate the contribution to $F(p,\eta_3, \eta_4)$ from the first region we can approximate $\left|\vec{q}_{1,2} - \vec{p}\right| \approx p$ and take into account that in the IR limit in question we have that $p\to 0$. Hence, we can Taylor expand around zero all the mode functions in the expression for $F(p,\eta_3, \eta_4)$. Then, the contribution to $F(p,\eta_3, \eta_4)$ from the first region, $\left|\vec{q}_{1,2}\right| \lesssim \left|\vec{p}\right|$, is:

\bqa
 && F_{(1)}(p,\eta_3,\eta_4) = \nonumber \\ && = \int \limits_{|q_1|,|q_2| < |p|} \frac{d^{D-1} q_1}{(2\pi)^{D-1}} \, \frac{d^{D-1} q_2}{(2\pi)^{D-1}} \, (\eta_3 \eta_4)^{D-2} h(q_1\eta_4)h^*(q_1 \eta_3)h(q_2\eta_4)h^*(q_2 \eta_3)h(p \eta_4)h^*(p\eta_3) \sim \nonumber \\ &&  \sim \int \limits_{|q_1|,|q_2| < |p|} d^{D-1} q_1 \,d^{D-1} q_2 \, q_1^{-2\nu} \,q_2^{-2\nu}\, p^{-2\nu}\,(\eta_3 \eta_4)^{D-2-3\nu} \sim (\eta_3 \eta_4)^{D-2-3\nu} p^{2(D-1-3\nu)}.
\eqa
In the second region (where  $q_{1,2} \geq p$) the largest IR contribution to $F(p,\eta_3, \eta_4)$ comes from $q_{1,2} \gg p$. Again we can Taylor expand all mode functions to obtain:

\bqa
  F_{(2)}(p,\eta_3,\eta_4) \sim \int \limits_{|q_1|,|q_2| > |p|} d^{D-1} q_1 \,d^{D-1}q_2 \, q_1^{-2\nu} \,q_2^{-2\nu} \,\left(\left|\vec{q}_1+\vec{q}_2\right|\right)^{-2\nu} (\eta_3 \eta_4)^{D-2-3\nu} \nonumber
\eqa
Finally, the estimates in the remaining two regions give the same result:

\bqa
F_{(3)}(p, \eta_3, \eta_4) \sim \int\limits_{\left|\vec{q}_1\right| < \left|\vec{p}\right|} d^{D-1} \vec{q}_1 \int\limits_{\left|\vec{p}\right| < \left|\vec{q}_2\right|} d^{D-1}\vec{q}_2 \, q_1^{-2\nu} \, q_2^{-4\nu} \, \left(\eta_3\eta_4\right)^{D-2-3\nu} \sim F_{(4)}(p,\eta_3, \eta_4).
\eqa
Therefore when $D-1-3\nu < 0$
there is a large IR contribution to $F(p,\eta_3, \eta_4)$
coming from the regions where either $q_1$ or $q_2$ or both are smaller than $p$.

Consequently in this case $n$ and $\kappa$ receive large IR contributions from  the integral over $q_{1,2}$ in the region $p < q_{1,2}$ and  the region $q_{1,2} < p$. This is one difference between the case of the complementary series and the principal series which gets large IR contributions only from the region $\left|\vec{q}_{1,2}\right| \gg \left|\vec{p}\right|$  \cite{AkhmedovPopovSlepukhin,Akhmedov:2013vka}.

Note however that  when $D-1-3\nu > 0$, $F(p,\eta_3, \eta_4)$ is well behaved
and $n_2$ and  $\kappa_2$  in (\ref{EPP}) receive large IR contributions only from $\left|\vec{p}\right| < \left|\vec{q}_{1,2}\right|$. I.e. in this case the situation is similar to the principal series \cite{Akhmedov:2013vka}. There is however a difference: in the principal series $n_2$, $\kappa_2$ receive logarithmic IR corrections \cite{Akhmedov:2013vka}
while here loop contributions are power like:
\bqa\label{leading}
n_2(p\eta) \approx \frac{\lambda^2}{3} |A_-|^2  \, (p\eta)^{-2\nu} \int \frac{d^{D-1}q_1}{(2\pi)^{D-1}}\frac{d^{D-1}q_2}{(2\pi)^{D-1}}\int \limits^1_\infty \int \limits^1_\infty d\eta_3 d\eta_4 (\eta_3 \eta_4)^{D-2-\nu}\times\nonumber\\\times
h(q_1 \eta_4)h^*(q_1 \eta_3) h(q_2 \eta_4) h^*(q_2 \eta_3)h(\left|\vec{q}_1+\vec{q}_2\right| \eta_4)h^*(\left|\vec{q}_1+\vec{q}_2\right| \eta_3), \nonumber \\
\kappa_2(p\eta) \approx -\frac{2\lambda^2}{3} \lambda^2(A_-^*)^2 \, (p\eta)^{-2\nu} \int \frac{d^{D-1}q_1}{(2\pi)^{D-1}}\frac{d^{D-1}q_2}{(2\pi)^{D-1}}\int \limits^1_\infty \int \limits^1_\infty d\eta_3 d\eta_4 (\eta_3 \eta_4)^{D-2-\nu}\times\nonumber\\\times
h(q_1 \eta_4)h^*(q_1 \eta_3) h(q_2 \eta_4) h^*(q_2 \eta_3)h(\left|\vec{q}_1+\vec{q}_2\right| \eta_4)h^*(\left|\vec{q}_1+\vec{q}_2\right| \eta_3), \nonumber \\
\eqa
These expressions are obtained from (\ref{EPP}) via the change of variables $q \to q\eta$, $\eta_{3,4} \to \eta_{3,4}/\eta$, neglecting $p$ in comparison with $q_{1,2}$ and expanding $h(p\eta_{3,4})$ to the leading order in the limit $p\eta_{3,4} \to 0$. But, after the substituting Eq. (\ref{leading}) into $\Delta_2 D^K$ in (\ref{EPP}) and expanding $h(p\eta_{1,2})$ around zero, these leading expressions cancel out. Moreover, the large IR contributions coming from the term $B\, x^{-\nu + 2}$ in the expansion of $h(x)$ also disappear from  the final expression for $\Delta_2 D^K$.

The largest IR correction to $\Delta_2 D^K$ comes from the subleading contributions to $n$ and $\kappa$.
To obtain it one has to express all the modes in Eq. (\ref{EPP}) using the Bessel functions  $J_\nu$ and $Y_\nu$ . Then in one of the four $h$'s $[h(p\eta_{1,2})$ and $h(p\eta_{3,4})]$ we have to single out $J_\nu\sim x^\nu$, as $x\to 0$, while in the other three --- $Y_\nu\sim x^{-\nu}$, as $x\to 0$. The corresponding expressions do not cancel out and provide the leading IR contribution to $\Delta_2 D^K$.

To calculate approximately the resulting leading contribution we neglect $p$ in comparison with $q_{1,2}$ under the integrals in (\ref{EPP}). There is however a change in the lower limits of integration over $\eta_{3}$ and $\eta_{4}$ which are set to  $\nu/p$. In doing this approximation we just neglect the contributions to $\Delta_2 D^K$ from the high physical momenta $p\eta_{3,4} \gg \nu$, where the physics is practically the same as in flat space.

After changing of the integration variables $u =  p \sqrt{\eta_3 \eta_4}$ and $v = \sqrt{\frac{\eta_3}{\eta_4}}$ we can harmlessly extend the integration over $v$ from infinity to zero. The integrals remain finite and the pre-factors of the expressions that we will find below, are just slightly changed by the contributions from the high physical momenta.

Finally , we expand $h(p\eta_{3,4})$ around zero, perform the integration over $u$ from $\nu$ to $p\eta$, and keep in the expression for $\Delta_2 D^K$ only the terms that are divergent/leading, as $p\eta \to 0$:

\bqa\label{DKBD}
\Delta_2 D^K(p|\eta_1,\eta_2) \approx \frac{8 \, A_-^3 A_+}{3(2\pi)^{2(D-1)}} \frac{\lambda^2 \, \log\left(p\eta/\nu\right) \, \eta^{D-1}}{(p\eta)^{2\nu}}\times \nonumber \\
\times \left\{\int \limits^\infty_1 dv \, v^{-D} \, G(v) \, \left(-\frac{1}{2\nu} v^{2\nu} + \frac{1}{v^{2\nu}}\right)-\int \limits^1_0 dv \, v^{-D} \, G(v) \, \left(\frac{1}{2\nu} v^{-2\nu} + v^{2\nu}\right)\right\},
\eqa
where

\bqa
G(v) = \int \int \frac{d^{D-1}q_1}{(2\pi)^{D-1}}\frac{d^{D-1}q_2}{(2\pi)^{D-1}} h(q_1 v^2) h^*(q_1) h(q_2 v^2) h^*(q_2)h(\left|\vec{q}_1+\vec{q}_2\right| v^2) h^*(\left|\vec{q}_1+\vec{q}_2\right|).
\eqa
Although $\lambda$ is small, the loop correction becomes comparable to the tree-level contribution. In fact, as $p \eta \to 0$, the sum of the tree--level and the second loop correction to the Keldysh propagator is as follows:

\bqa\label{dplusdelta}
D^K_{0} + \Delta_2 D^K \approx \eta^{D-1}/(p\eta)^{2\nu}\left[a + b \, \lambda^2 \, \log \frac{p\eta}{\nu}\right],
\eqa
where the constants $a$ and $b$ are computed from expressions above.

As a side remark, if $D - 1 - 3\,\nu < 0$,  the IR contributions to $n_2$ and $\kappa_2$  have the following form:
\bqa\label{17}
n_2(p\eta) \propto \lambda^2 (p\eta)^{D-1-6\nu}, \quad
\kappa_2(p\eta) \propto \lambda^2 (p\eta)^{D-1-6\nu}. \quad
\eqa
We do not present here the full expressions for $n_2$, $\kappa_2$ and $\Delta_2 D^K$ because in this case we will not do the resummation of the leading IR contributions from all loops.
Again, after  substituting the above  leading expressions (\ref{17}) into $\Delta_2 D^K$ they cancel out. What survives in $\Delta_2 D^K$ is coming from subleading contributions to $n_2$, $\kappa_2$ and $\kappa_2^*$. Corrections to $\Delta_2 D^K$ are always logarithmic as in (\ref{DKBD}) and (\ref{dplusdelta}), but the coefficients in front of them are different depending on the case when $D-1-3\,\nu$ is greater or lower  than zero.

\subsection{Correction to the vertices}
\label{IIIB}
 For the vertices it is more convenient to use the non-stationary diagrammatic technique before the Keldysh rotation (see e.g. \cite{Akhmedov:2013vka} for the explanation and notations). Then the one--loop correction from the $(--)$ ``fish'' diagrams to the vertices is as follows:

\bqa\label{vert12}
\lambda^{--}(\eta_1,\eta_2,p_1,p_2,p_3,p_4) = (-i \lambda)^2 (\eta_1 \eta_2)^{D-1}\delta^{(D-1)}\left(\vec{p}_1 + \vec{p}_2 + \vec{p}_3+\vec{p}_4\right) \int \frac{d^{D-1} q}{(2\pi)^{D-1}}\times \nonumber \\ \times
\left\{\theta(\eta_1-\eta_2)^{\phantom{\frac12}} h(\left|\vec{q}-\vec{p_1}\right|\eta_1) h^*(\left|\vec{q}-\vec{p_1}\right|\eta_2) + \theta(\eta_2-\eta_1)\,\, h(\left|\vec{q}-\vec{p_1}\right|\eta_2) h^*(\left|\vec{q}-\vec{p_1}\right|\eta_1)\right\}\times \nonumber \\
\times \left\{\theta(\eta_2-\eta_1)^{\phantom{\frac12}} h\left(\left|\vec{p}_2+ \vec{q}\right|\eta_2\right) h^*\left(\left| \vec{p}_2 + \vec{q}\right|\eta_1\right) + \theta(\eta_1-\eta_2)\,\, h\left(\left|\vec{p}_2 + \vec{q} \right|\eta_1\right) h^*\left(\left|\vec{p}_2 + \vec{q}\right|\eta_2\right)\right\}
\eqa
Here the indexes ``$+$'' and ``$-$'' are attributed to the two internal vertices in the one loop diagram describing correction to the tree-level vertex.
The situation with the other vertices, $\lambda^{+-}$, $\lambda^{-+}$ and $\lambda^{++}$ is very similar. We would like to check if (\ref{vert12}) contains large corrections in the limit $p_i\eta_{1,2}\to 0$, $i=1,2,3,4$.

As in the case of the Keldysh propagator, we divide the domain of the integration over the internal momentum $\vec{q}$ into regions. One region is where $q \geq (p_1 p_2 p_3 p_4)^{\frac14}$. To estimate $\lambda^{--}$ in this domain we observe that the largest contribution to the vertices comes from $q \sim \left(p_1 p_2 p_3 p_4\right)^{1/4} \to 0$. (In the next section we present more rigorous observations.) Hence, Taylor expanding all mode functions around zero, we obtain that the contribution to the vertex from the first region is as follows:

\bqa\label{IR2}
&& \Delta_1 \lambda^{--}(\eta_1,\eta_2,p_1,p_2,p_3,p_4) \sim \nonumber \\ && \sim (-i \lambda)^2 (\eta_1 \eta_2)^{D-1-2\nu} \, \delta^{(D-1)}\left(\vec{p}_1 + \vec{p}_2 + \vec{p}_3 + \vec{p}_4\right) \,
\int\limits_{|q| > \sqrt[4]{p_1 p_2 p_3 p_4}} d^{D-1} q \, q^{-4\nu}. \label{vertex}
\eqa
If $D-1-4\nu<0$, this expression is large. In the opposite case, when $D-1-4\nu>0$, the integral in (\ref{vertex}) is convergent in the IR limit under consideration.

To estimate the vertex contribution in the domain of integration  where $q \leq (p_1 p_2 p_3 p_4)^{\frac14}$, we can neglect $q$ in comparison with $p_i$. Then

\bqa
&& \Delta_2 \lambda^{--}(\eta_1,\eta_2,p_1,p_2,p_3,p_4) \sim \nonumber \\ && \sim  (-i \lambda)^2 (\eta_1 \eta_2)^{D-1-2\nu} (p_1 p_2)^{-2\nu} \, \delta^{(D-1)}\left(\vec{p}_1 + \vec{p}_2 + \vec{p}_3 + \vec{p}_4\right) \, \int\limits_{|q| < \sqrt[4]{p_1 p_2 p_3 p_4}} d^{D-1} q.
\eqa
Again, if $D-1-4\nu>0$, this expression does not contain large contributions. If, however, $D-1-4\nu < 0$, there are large IR corrections, which are similar to (\ref{IR2}). In the next subsection we discuss more rigorously the physical origin and the meaning of these problems with the vertex corrections.

\section{The physical roots of the secular effects in the vertexes and self--energies}
\label{IV}
Thus, when
\begin{equation}
\nu >\frac{D - 1}4
\end{equation}
(i.e. when $m < \frac{\sqrt{3}}{4}(D-1)$)  there are potentially dangerous IR corrections to the vertices, as $p_i\eta_i \to 0$. Furthermore, when
\begin{equation}
\nu >\frac{D - 1}3
\end{equation}
(i.e. when $m < \frac{\sqrt{5}}{6}(D-1)$) there are also potentially dangerous IR contributions to the Keldysh propagator (self--energy). Such contributions can complicate the problem of the summation of the leading IR corrections in all loops because  the IR limit of the entire system of the Dyson--Schwinger equations has to be solved for propagators and vertices simultaneously. This is to be compared to the case of the principal series, where the problem may be reduced to the solution of only one Dyson--Schwinger equation for the Keldysh propagator \cite{Akhmedov:2013vka}.


To better understand what is going on recall that the spectrum of a theory of a neutral scalar meson in flat space is made of the mass shell $p^2=m^2$ plus a continuum including all two-particle states, the three-particle states etc. which starts at $p^2 = 4m^2$. The very existence of the complementary series of fields in dS space makes the situation rather different. The situation becomes clearer when we express, say, the BD two--point functions by using the coordinate independent plane wave representation introduced in \cite{bgm,Bros}.

To this aim, in this section  we look at either the real or the complex dS manifold as submanifolds of the complex Minkowski manifold with one
spacelike dimension more:
\beq
dS_D = \{x \in M_{D+1}\ :\ x\cdot x = -R^2=-1\}\ \ \ {\rm and}\ \ \
dS_D^{(c)} = \{z \in M_{D+1}^{(c)}\ :\ z\cdot z = -R^2=-1\}\
\label{s.2}\endq
where
$
x\cdot y =
x^0y^0- \vec{x}\cdot\vec{y}.
$
The dS metric is obtained by restriction of
the ambient space-time interval to $dS_D$.
In particular, for  the EPP  we get
\begin{eqnarray}
&&x(t,\x) = \ \left\{
\begin{array}{lcl}
x^0 &=&  \sinh t +  \frac{e^t}{2}|\x|^2  \\
x^i &=& {e^t}\x^i \\
x^{D} &=&\cosh t -  \frac{e^t}{2}|\x|^2  \\
\end{array}  \right.
\label{s.4}\end{eqnarray}
\begin{equation}
ds^2= (dx_0^2-dx_1^2-\ldots dx_D^2)|_{dS_D} = dt^2 - \exp(2 t)\ d\x^2.
\end{equation}
There exists a remarkable set of solutions of the dS Klein--Gordon equation which
may be interpreted as dS plane waves \cite{bgm, Bros,faraut,molcanov}.
For a complex dS event $z$, a given  nonzero lightlike vector $\xi\in C_+$ and a complex number $\lambda \in \bC$   the homogeneous function
\begin{equation}
z\mapsto (\xi\cdot z)^\lambda
\label{pwholo}\end{equation}
 satisfies there  the massive (complex) Klein-Gordon equation:
\begin{equation}
(\Box_{ z}+ m_\lambda^2)(\xi\cdot z)^\lambda =0. \label{cdskg}
\end{equation}
The above plane waves are holomorphic in the
future and past dS tuboids $\TT_\pm$ (that are related to the spectral condition of dS QFT \cite{bgm,Bros,bem}) obtained as the intersections of the ambient tubes \cite{sw} with the complex dS manifold:
\beq
\TT_\pm = \{x\pm iy \in dS_d^{(c)}\ :\ y^2= y\cdot y >0, \ \  y^{0}>0 \}\ .
\label{s.2.1}\endq
The parameter $\lambda$ is here unrestricted, i.e. we may consider
{\em complex squared masses} $m^2_\lambda = - \lambda(\lambda+D-1)$.
The symmetry  $\lambda \to (-\lambda-D+1)$ also implies that
\beq
\left(\Box_z +m_\lambda^2\right)(\xi\cdot z)^{-\lambda-D+1} =0.
\label{p.2abis}
\endq
The two boundary values of the complex waves (one from each tuboid)
are homogeneous distributions of degree $\lambda$
and are solutions of the dS Klein--Gordon equation {\em on the real dS manifold}:
\beq
\left(\Box_x +m_\lambda^2\right)(\xi\cdot x)_\pm^\lambda =0, \ \ \ \ \ \ \ \  \left(\Box_x +m_\lambda^2\right)(\xi\cdot x)_\pm^{1-D-\lambda} =0.
\label{p.2}\endq
The waves depend in a $\CC^\infty$ way on $\xi$ and are entire in $\lambda$.

We may now introduce a class of maximally analytic vacua (the BD vacua --- one for each complex squared mass) by specifying their two-point functions. In the following plane-wave expansion we take the first point $z$ in the backward tuboid
$\TT_-$, the second point $z'$ in the forward tuboid ${\TT}_+$ and $\lambda$ is not a pole of $\Gamma(-\lambda)\Gamma(\lambda+D-1)$:
\begin{eqnarray}
&& \Wl({z}, {z'}) = \wl({z}\cdot {z'}) = {\Gamma(-\lambda)\Gamma(\lambda+D-1)
e^{-i\pi\left ( \lambda+{D-1\over 2} \right )}\over 2^{D+1}\pi^D}
\int_{S_0}(\xi\cdot {z})^{1-D-\lambda}(\xi\cdot {z'})^{\lambda}\,
d\xx.
\label{r.3}
\end{eqnarray}
The integral is performed  on the spherical basis $S_0$ of the future cone: $S_0= \{\xi:\ \xi^2=0, \   \xi_0=1\}$; $d\xx$ denotes the spherical invariant measure. One gets that the above Fourier-like representation may be evaluated in terms of Legendre functions of the first kind:
\begin{align}
\wl({z}\cdot {z'}) =
{\Gamma(-\lambda)\Gamma(\lambda+D-1)\over 2 (2\pi)^{D/2}}
(\zeta^2-1)^{-{D-2 \over 4}}\,P_{\lambda +{D-2\over 2}}^{-{D-2 \over 2}}(\zeta),
\ \ \ \zeta = {z}\cdot {z'}\ .
\label{r.1}\end{align}
It is clear from (\ref{r.1}) that
$\zeta \mapsto \wl(\zeta) = w_{-\lambda-D+1}(\zeta)$
is holomorphic in $\bC\setminus (-\infty,\ -1]$ i.e. everywhere except on the locality cut ({\em maximal analyticity property}).

When we specialize the above construction to fields having real and positive squared masses
we immediately understand that there are two types of waves: the {principal series} $\lambda= -\frac{D-1}2+ i \nu$ with $\nu\in {\bf R}$;
in this case waves have an oscillatory character:
\begin{equation}
\phi(x)= (\xi\cdot x)_\pm^{-\frac{D-1}2+ i \nu},  \ \ \ \ \ \ \ \ \ m^2 =  \left(\frac{D-1}2\right)^2+  \nu^2.
\end{equation}
The {complementary series} $\lambda= -\frac{D-1}2+ \nu $ with $|\nu| < (D-1)/2$.
Here waves  do not oscillate and decay more slowly at infinity:
\begin{equation}
 \phi(x)= (\xi\cdot x)_\pm^{-\frac{D-1}2 + \nu},
 \ \ \ \ \ \ \ \ \  m^2 =  \left(\frac{D-1}2\right)^2 -   \nu^2.
 \end{equation}
Although redundant, let us write explicitly the above Fourier-like representation of the BD vacua in the two cases of interest: for $\nu\in {\bf R}$ theories of the principal series have the following two-point functions:
\begin{eqnarray}
&& W_{i\nu}({z}, {z'}) =w_{i\nu}(\zeta) = {\Gamma\left(\frac{D-1}2 + i \nu\right)\Gamma\left(\frac{D-1}2 - i \nu\right)
\over 2^{D+1}\pi^D e^{-\pi\nu}}
\int_{S_0}(\xi\cdot {z})^{-\frac{D-1}2 - i \nu}(\xi\cdot {z'})^{-\frac{D-1}2 + i \nu}
d\xx.
\label{princ}
\end{eqnarray}
For $\nu\in {\bf R}$ and  $ |\nu| < (D-1)/2$ theories of the complementary series have the following two-point functions:
\begin{eqnarray}
&& W_{\nu}({z}, {z'})  = w_{\nu}(\zeta)= {\Gamma\left(\frac{d-1}2 + \nu\right)\Gamma\left(\frac{D-1}2 - \nu\right)
\over 2^{D+1}\pi^D e^{i\pi\nu}}\
\int_{S_0}(\xi\cdot {z})^{-\frac{D-1}2 -  \nu}(\xi\cdot {z'})^{-\frac{D-1}2 + \nu}
d\xx.
\end{eqnarray}
Now let us come back to our main line of thought. When studying the corrections to the propagators and to the vertices we are led to consider the distributions $W^2(x,x')$ and $W^3(x,x')$.
Let us focus on the Wick-square $W^2(x,x')$ of the two-point function and consider a theory of the principal series. The asymptotic behaviour can be crudely estimated by looking at the square of a plane wave: $(\xi\cdot x)_\pm^{-(D-1)+ 2 i \nu}$.
This function behaves at infinity  better than a wave of the principal series; since $W^2_{i\nu}(x,x')$  is a positive-definite distribution, it should be possible to write an expansion of it just in terms of two-point functions of the principal series as follows:
\begin{equation}
W^2_{i\nu}(\zeta) = \int \kappa \rho_{i\nu}(i\kappa) W_{i\kappa}(\zeta) d\kappa.
\end{equation}
This property should remain true also for fields of the complementary series as long as
\begin{equation}
-(D-1)+ 2  |\nu| \leq- \frac{D-1}2 \label{bound}
\end{equation}
i.e. as long as $|\nu|\leq \frac{D-1}4$. To understand what happens when this bound is violated we need to examine the above  K\"all\'en--Lehmann representation more closely. The problem of finding the weight $\rho$ has been solved in \cite{Bros:2009bz} in a more general case, namely for the product of two distributions  $w_{i\nu}(\zeta)$ and $ w_{i\lambda}(\zeta)$ belonging to the principal series (\ref{princ}); there holds the  following integral representation:
\beq
w_{i\nu}(\zeta) w_{i\lambda}(\zeta)  =
\int_\bR \kappa \rho_{i\nu,i\lambda}(i\kappa)w_{i\kappa}(\zeta)\,d\kappa\
\label{c.1}\endq
where the K\"all\'en--Lehmann weight has the following remarkable explicit  expression:
\beqa
{
\kappa\rho_{i\nu,i\lambda}(i\kappa) =
{1\over 2^5 \pi^{D+5\over 2}
\Gamma\left ( {D-1\over2} \right )}
{\kappa\,\sinh \pi\kappa \,
\prod_{\epsilon, \epsilon',\epsilon'' = \pm 1}
\Gamma \left({D-1\over 4}
+{i\epsilon\kappa +i\epsilon'\nu +i\epsilon''\lambda \over 2}
\right ) \over
\Gamma\left ( {D-1\over2} +i\kappa \right )\Gamma\left ( {D-1\over2} -i\kappa \right )}} \endqa
By using  such explicit result we may perform analytic continuation in the mass parameters to obtain the K\"all\'en-Lehmann representation  for the product  $w_{\alpha}(\zeta)\,w_{\beta}(\zeta)$
belonging to the complementary series and violating the bound $\alpha+\beta<\frac{D - 1}2$ (see \cite{Bros:2009bz} for more details). The result is as follows: if $N$ is a non-negative integer such
that
\begin{equation}
 \frac{D - 1}4  +N < \frac 12 (\alpha +  \beta) < \frac{D - 1}4+N+1
 \end{equation}
provided
\begin{equation}
N <  \frac{D - 1}4
 \end{equation}
 the K\"all\'en--Lehmann representation of the product $w_{\alpha}(\zeta)\,w_{\beta}(\zeta)$ incudes $N+1$ discrete terms:
\beqa
{
w_{\alpha}(\zeta)\,w_{\beta}(\zeta) =
\int_\bR \n\,\rho_{\alpha,\beta}(i\n)\,w_{i\n}(\zeta)\,d\n}
\nonumber+ \sum_{n=0}^N A_n(\alpha,\ \beta)\,w_{\frac{D - 1 -2\alpha-2\beta}2+2n}(\zeta)
\label{u.22}\endqa
where
\beqa
&&{A_n(\alpha,\beta) =
{\Gamma(\alpha-n)\Gamma(\beta-n) \Gamma(\alpha+\beta-n)
\Gamma(\frac{D - 1}2-\alpha+n)\Gamma(\frac{D - 1}2-\beta+n)\Gamma(\frac{D - 1}2+n)\over
4 n!\pi^{D+1\over 2}\Gamma\left ( {D-1\over2} \right )
\Gamma(-\frac{D - 1}2-2n+\alpha+\beta)\Gamma(-2n+\alpha+\beta)
\Gamma({D - 1}+2n-\alpha-\beta)} \times}\cr &&\times
{(-1)^n \Gamma(n+\frac{D - 1}2-\alpha-\beta)\over
\Gamma(2n+\frac{D - 1}2-\alpha-\beta)}\
\label{u.22.1}\endqa
provided neither $\alpha$ nor $\beta$ is an integer (if
$\alpha + \beta < \frac{D-1}2$ the formula holds without the $A_n$ terms as for the principal series).
It is easy to check that $\n\,\rho_{\alpha,\beta}(\n) \ge 0$. Furthermore
all the factors in $A_n(\alpha,\beta)$ -  except the last fraction - are positive
since the arguments of the Gamma functions are positive.
The last fraction is of the form
\beq
{(-1)^n \Gamma(n+x)\over \Gamma(2n+x)} = (-1)^n \prod_{q}^{2n-1} (q+x)^{-1}\ .
\label{u.22.2}\endq
The last product contains $n$ negative factors and the result is positive.

The conclusion that can be drawn from the above analysis may be surprising.
Let us consider to fix the ideas a dS space--time dimension $D=4$ and consider free fields of the complementary series violating the bound (\ref{bound}), i.e.  fields such that $\nu>3/4$ for the complementary series.
From the above analysis it follows that the two-particle subspace of the Fock space relative to a field like that contains a {\em discrete component} of parameter $2\nu-3/2$.
In other words a free field theory of mass
\begin{equation}
m^2 = \frac 9 4 - \nu^2 < \frac {27} {16}
\end{equation}
contains in the two-particle states discrete terms ("bound states") of mass
\begin{equation}
M^2 = (3 - 2\nu) 2\nu < \frac {9} {4}
\end{equation}
Similarly, when considering the  the three-particle subspace of a light free field, one sees that the behaviour at infinity is not worse that hat of the principal series provided
\begin{equation}
-\frac 3 2 (D-1)+ 3  \nu \leq- \frac{D-1}2 \label{bound2}
\end{equation}
On the other hand, a more laborious calculation show that when $\nu> \frac{D-1}3$  as before  "bound states" appear in the three particle subspace of the theory.

\section{Remarks on loop corrections for the out-modes}

To  complete discussion we have also to consider loop corrections for other $\alpha$--modes. In fact, we have here an interacting theory and take it in the IR limit. Then, modes different from the BD modes may play an important role: for instance for the principal series it happens that due to interactions the in--ground state is transformed in the out--ground state in future infinity \cite{Akhmedov:2013vka} and, hence, the loop resummation has to be done with the use of out--modes. On the other hand it is not hard to see that the structure of the loop corrections for a generic $\alpha$--modes is very similar to the BD case. The only exceptions are the out--modes in a sense that we explain now.

For the out--modes $h(x) \propto J_\nu(x)$ while $h^*(x) \propto i\, Y_\nu(x)$. These modes are related to the BD-modes via a Bogolubov rotation. The main difference in comparison with the BD modes is that for $x\to 0$ the behaviour of the modes is $h(x) \approx B_+ x^{\nu}$  and $h^*(x) \approx B_- x^{-\nu}$, as $x \to 0$; here $B_+$ is some real constant, while $B_-$ is purely imaginary;  $h^*$ is not the complex conjugate of $h$\footnote{Note that in the case of the principal series, when $m > (D-1)/2$, $\nu$ is imaginary. Hence, in our previous papers we were able to choose $J_\nu$ and its complex conjugate as the basis of mode functions. However, when $\nu$ is real, then $J_\nu$ is also real. As the result, we have to choose $J_\nu$ and $i \, Y_\nu$ as the basis. This way we obtain the proper commutation relations for the creation and annihilation operators. Similar situation one encounters in flat space, if does the Bogolubov rotation from $e^{\pm i \omega t}$ modes to $\cos(\omega t)$ and $i\,\sin(\omega t)$.}. Thus, for the out--modes the IR behavior of $h^*$ is different from that of $h$ contrary to what happens for BD modes and even for generic $\alpha$--modes.
Note however that in the UV limit the out--modes behave as $h(p\eta)\sim \frac{\cos(p\eta - \nu\pi/2 - \pi/4)}{\sqrt{p\eta}}$ and $h^*(p\eta)\sim \frac{\sin(p\eta - \nu\pi/2 - \pi/4)}{\sqrt{p\eta}}$. These mixtures of the positive and negative energy modes $e^{\pm i \, p \, \eta}$ spoil the UV behavior of the corresponding propagators. This is the characteristic feature of a generic $\alpha$--mode, the only exception being the BD modes, which exhibit the standard UV behavior. Note, however, that although out-modes have an incorrect UV behavior  they can be relevant in the IR limit in presence of an interaction (the prototypical example is the Cooper pairing that is incorrect in UV limit but provides a correct description of the IR physics).

For the out-modes the two-loop large IR correction to the Keldysh propagator is also contained in the expressions of Eq. (\ref{EPP}). It is not hard to show that for any mass parameter $\nu$ the function $F(p,\eta_3,\eta_4)$  contains no  large IR contribution. The calculation of $n_2$, $\kappa_2$ and $\kappa_2^*$ proceeds along the same lines as for the principal series in  the BD case \cite{Akhmedov:2013vka}. The final answer for $n_2$, $\kappa_2$ and $\kappa_2^*$ is the following:
\bqa
n_2(p\eta) \approx \frac{2 \lambda^2 B_+ B_-}{3}\log\left(\frac{\nu}{p\eta}\right)\int \frac{d^{D-1} l_1}{(2\pi)^{D-1}}\frac{d^{D-1} l_2}{(2\pi)^{D-1}} \int^\infty_0 dv \nonumber\times \\ \times
v^{2 D - 2\nu -3}\, h(l_1 \, v^2)\,h^*(l_1)\, h(l_2 \, v^2)\,h^*(l_2)\, h\left(\left|\vec{l}_1+\vec{l}_2\right| \, v^2\right)\,h^*\left(\left|\vec{l}_1+\vec{l}_2\right|\right), \nonumber\\
\kappa_2(p\eta) \approx - \frac{2\lambda^2 B_-^2}{3 \, \nu} (p \eta)^{-2\nu} \, \int \frac{d^{D-1} l_1}{(2\pi)^{D-1}}\frac{d^{D-1} l_2}{(2\pi)^{D-1}} \int^1_0 dv \nonumber\times \\\times
v^{2 D - 2\nu -3}\, h(l_1 \, v^2)\,h^*(l_1)\, h(l_2 \, v^2)\,h^*(l_2)\, h\left(\left|\vec{l}_1+\vec{l}_2\right| \, v^2\right)\,h^*\left(\left|\vec{l}_1+\vec{l}_2\right|\right), \nonumber\\
\kappa_2^*(p\eta) \approx - \frac{2\lambda^2 B_+^2}{3\, \nu} (p \eta)^{2\nu} \, \int \frac{d^{D-1} l_1}{(2\pi)^{D-1}}\frac{d^{D-1} l_2}{(2\pi)^{D-1}} \int^1_0 dv \nonumber\times \\\times
v^{2 D - 2\nu -3}\, h(l_1 \, v^2)\,h^*(l_1)\, h(l_2 \, v^2)\,h^*(l_2)\, h\left(\left|\vec{l}_1+\vec{l}_2\right| \, v^2\right) \, h^*\left(\left|\vec{l}_1+\vec{l}_2\right|\right).
\eqa
Taking into account the behavior of $h(p\eta_{1,2})$ and $h^*(p\eta_{1,2})$ at future infinity we find that the leading IR correction to the Keldysh propagator comes from $n$ alone and is logarithmic:
\bqa
\Delta_2 D^K(p,\eta_1,\eta_2) = \eta^{D-1} \, B_+B_-\left[ \left(\frac{\eta_1}{\eta_2}\right)^\nu+ \left(\frac{\eta_2}{\eta_1}\right)^\nu\right] \, n_2(p\eta).
\eqa
It seems that out--modes may allow the loop resummation for light fields $D - 1 - 4\nu < 0$ as is the case for the principal series \cite{Akhmedov:2013vka}. The point is that the resummation can be done by solving the system of Dyson--Schwinger equations, but this system is covariant under the simultaneous Bogolubov transformation of the modes and $n$, $\kappa$ and $\kappa^*$. Moreover, it is straightforward to show that for the out--modes the vertices do not receive any large corrections in the limit $p_i\eta_{1,2} \to 0$, $i=1,2,3,4$. This is true because of the peculiar relation between the IR behavior of $h$ and $h^*$ for the out-modes.

Thus, it seems convenient to try and solve the system of the Dyson--Schwinger equations with the use of out--modes. However, unlike the case of the principal series \cite{Akhmedov:2013vka}, any small excitation of $\kappa$ and $\kappa^*$  on top of the out--ground state leads to growing rather than damping effects. This can be seen in the  Dyson--Schwinger system of equations at linear order in $\kappa$ and $\kappa^*$ (see \cite{Akhmedov:2013vka} for the details and methods).

\section{Resummation: preliminary discussion}

Thus, for all the $\alpha$--modes  loop effects ($\sim \lambda^2 \log(p\eta)$) can become large as $p\eta \to 0$ even when $\lambda^2$ is very small. 
The important point is that loop corrections are not suppressed in comparison with classical tree level contributions to propagators and vertices. Hence, to understand the physics in dS space, one has to sum unsuppressed IR corrections from all loops.

In the EPP there is also the distinct problem of summing the dS-invariant IR corrections to the correlation functions of the exact BD state (see, e.g., \cite{Gautier:2015pca}--\cite{Serreau:2013psa} and \cite{AkhmedovBurda}, \cite{Youssef:2013by}). Only in this case  loop corrections respect the dS isometry (in the EPP) \cite{Polyakov:2012uc} (see also \cite{AkhmedovBurda}, \cite{Akhmedov:2013vka} and \cite{Higuchi:2010xt}). Here however we propose to consider an initial nonsymmetric density perturbation on top of the BD state at  past infinity of the EPP. We would like to trace the destiny of such a perturbation and to understand the effect of the large IR contributions, as the system progresses towards future infinity.

As we have explained in the introduction  we cannot just put initial comoving density $n^0_p$ at past infinity of the EPP, because then the physical density will be infinite. One has to put the initial value $n^0_p$ at an initial Cauchy surface $+\infty > \eta_0 > 0$. Moreover, due to the UV divergences the comoving momentum also should be cutoff at the UV scale $p_0$. But, as was explained above $n(p\eta)$ and $\kappa(p\eta)$ are attributed to the comoving volume and, hence, do not change before $p\eta \sim \nu$. Their behaviour for $p\eta > \nu$ is not much different from that in flat space--time. Thus, cutting simultaneously comoving momentum and conformal time integrals effectively amounts to cut the physical momentum integrals at $\nu$. On the other hand, due to the symmetries of the EPP (or isometries of the dS space) we can put an initial comoving density at an initial value of the physical momentum $P_0 \equiv (p\eta)_0 \sim \nu$ and cutoff all the integrals over the physical momentum at this value. That is what was proposed in the introduction.

To sum loop contributions we should solve the  Dyson--Schwinger system of equations for the propagators, self-energies and vertices. We would like to sum only powers of the leading contribution $\lambda^2 \log(p\eta)$ and neglect subdominant terms, such as powers of $\lambda^4 \log(p\eta)$ or $\lambda^2 \log(\eta_1/\eta_2)$ etc..

As we have seen above, the retarded and advanced propagators do not receive large IR contributions from the first and second loops, but only from higher loops which may be produced from the lower loop corrections to the Keldysh propagator. This means that such contributions to $D^{R,A}$ are suppressed by higher powers of $\lambda$. Since we want to sum only the leading corrections we may use everywhere the tree level expressions for  $D^{R,A}$ (renormalized with the leading UV corrections). That is what we will always do below.

However,  at variance with the principal series case \cite{Akhmedov:2013vka}, we saw that in the complementary case large IR contributions to the vertices may also arise. This happens for generic $\alpha$--modes. To avoid that difficulty we restrict here  our consideration to fields with $m$ sufficiently large ($D-1-4\nu > 0$) $m > \frac{\sqrt{3}}{4} (D-1)$.

When the resummation is done for the exact BD state in the  EPP,  the leading contributions come from the summation of the sunset bubbles. In fact, if one puts the above two loop logarithmic correction to the Keldysh propagator into the internal legs of the sunset diagram, the correction is suppressed as $\lambda^4 \, \log(p\eta)$ (because of its logarithmic behavior and of the integration over the complete range of the physical momentum inside the loops). The situation is very similar to the standard UV renormalization: if one puts loop corrected expressions again inside the loops they lead to subleading corrections, while the leading corrections come from the multiplication of the bubbles. That is exactly the reason why in the Dyson--Schwinger equation one can put the exact Keldysh propagator only into one of the external legs. This fact was used in part in \cite{Gautier:2015pca}, \cite{Gautier:2013aoa}, \cite{Serreau:2013psa}. As a result in this case the Dyson--Schwinger equation reduces to a linear integro--differential equation. However, if we cut the integration over the physical momentum at $P_0$  and put an initial value $n\left(P_0\right)$, for the comoving density of the exact modes, the situation becomes very different. Now the internal legs also bring leading corrections of the type $\left|\lambda^2 \, \log(p\eta)\right|^n$. Then one has to put the exact Keldysh propagators also in the internal legs inside the loops and there follows a non--linear integro--differential equation. The latter one has much reacher realm of solutions.

\section{IR solution of the Dyson--Schwinger equations}

In the previous section  we have justified  why  for BD--modes and  for $D-1-4\nu > 0$
the  Dyson--Schwinger system of equations reduces to the equation for the Keldysh propagator alone:

\bqa\label{DSDK2}
D^{K}(p|\eta_1,\eta_2) \approx D_{0}^K(p|\eta_1,\eta_2) +\frac{\lambda^2}{6} \int \frac{d^{D-1}\vec{q_1}}{(2\pi)^{D-1}}\frac{d^{D-1}\vec{q_2}}{(2\pi)^{D-1}} \iint_{+\infty}^0 \frac{d\eta_3 d\eta_4}{(\eta_3\eta_4)^D} \times \nonumber \\ \times \Biggl[ \Biggr.
3\, D^K_0 \, \left(p\,|\eta_1,\eta_3\right) \, D^K\left(q_{1}|\eta_3,\eta_4\right) \, D^K\left(q_{2}\, |\eta_3,\eta_4\right) \,D_0^A\left(\left|\vec{p}-\vec{q}_1-\vec{q}_2\right|\, |\eta_3,\eta_4\right) \, D_0^A\left(p\, |\eta_4,\eta_2\right)
- \nonumber \\-
\frac{1}{4}\,D^K_0 \, \left(p\,|\eta_1,\eta_3\right) \, D_0^A\left(q_{1}|\eta_3,\eta_4\right) \, D_0^A\left(q_{2}\, |\eta_3,\eta_4\right) \,D_0^A\left(\left|\vec{p}-\vec{q}_1-\vec{q}_2\right|\, |\eta_3,\eta_4\right) \, D_0^A\left(p\, |\eta_4,\eta_2\right)
- \nonumber \\-
\frac{3}{4}\,D_{0}^R\left(p\,|\eta_1,\eta_3\right) \, D^K\left(q_{1}|\eta_3,\eta_4\right) \, D_0^A\left(q_{2}\, |\eta_3,\eta_4\right) \,D_0^A\left(\left|\vec{p}-\vec{q}_1-\vec{q}_2\right|\, |\eta_3,\eta_4\right) \, D_0^A\left(p\, |\eta_4,\eta_2\right)
+ \nonumber \\+
D_{0}^R\left(p\,|\eta_1,\eta_3\right) \, D^K\left(q_{1}|\eta_3,\eta_4\right) \, D^K\left(q_{2}\, |\eta_3,\eta_4\right) \,D^K\left(\left|\vec{p}-\vec{q}_1-\vec{q}_2\right|\, |\eta_3,\eta_4\right) \, D_0^A\left(p\, |\eta_4,\eta_2\right)
- \nonumber \\-
\frac{3}{4}\,D_{0}^R\left(p\,|\eta_1,\eta_3\right) \, D^K\left(q_{1}|\eta_3,\eta_4\right) \, D_0^R\left(q_{2}\, |\eta_3,\eta_4\right) \,D_0^R\left(\left|\vec{p}-\vec{q}_1-\vec{q}_2\right|\, |\eta_3,\eta_4\right) \, D_0^A\left(p\, |\eta_4,\eta_2\right)
- \nonumber \\-
\frac{1}{4}\,D_{0}^R\left(p\,|\eta_1,\eta_3\right) \, D_0^R\left(q_{1}|\eta_3,\eta_4\right) \, D_0^R\left(q_{2}\, |\eta_3,\eta_4\right) \,D_0^R\left(\left|\vec{p}-\vec{q}_1-\vec{q}_2\right|\, |\eta_3,\eta_4\right) \, D^K\left(p\, |\eta_4,\eta_2\right)
+ \nonumber \\+
3\,D_{0}^R\left(p\,|\eta_1,\eta_3\right) \, D_0^R\left(q_{1}|\eta_3,\eta_4\right) \, D^K\left(q_{2}\, |\eta_3,\eta_4\right) \,D^K\left(\left|\vec{p}-\vec{q}_1-\vec{q}_2\right|\, |\eta_3,\eta_4\right) \, D^K\left(p\, |\eta_4,\eta_2\right)
\Biggl. \Biggr].
\eqa
This equation is formally the same as the previous Eq. (\ref{DSDK}). The difference is that at its RHS appears wherever appropriate the exact propagator  $D^K$ and tree--level propagators $D_0^R$ and $D_0^A$. Again, equation (\ref{DSDK2}) is invariant under simultaneous Bogolubov rotations of the modes and consequently of $n$, $\kappa$ and $\kappa^*$.

We want to solve this equation in the infrared limit $p\eta \lesssim \nu$ where modes behave as $h(p\eta) = A_- \left(p\eta\right)^{-\nu} + i A_+ \left(p\eta\right)^{\nu}$. For the reasons  mentioned at the beginning of the section III we restrict our attention to the case $\nu \leq 1$. For $D\leq 5$ this includes the above restriction on $\nu$. In the IR limit under consideration we obtain

\bqa\label{DRAIR}
D_{0}^{\frac{R}{A}}\left(p\,|\eta_1,\eta_2\right) = \mp \theta\left(\mp\Delta\eta_{12}\right) \,2 A_- A_+ \,(\eta_1\eta_2)^{\frac{D-1}{2}}\,\left[\left(\frac{\eta_1}{\eta_2}\right)^{\nu}-\left(\frac{\eta_2}{\eta_1}\right)^{\nu}\right],\quad p\eta_{1,2} \to 0.
\eqa
We use the same ansatz for the Keldysh propagator that was introduced in a previous papers for the principal series \cite{Akhmedov:2013vka}:

\bqa\label{ansatz0}
D^{K}\left(p\, |\eta_1,\eta_2\right) \approx \eta^{D-1}\, \left\{h(p\eta_1)\, h^*(p\eta_2) \,\left[\frac12 + n_e(p\eta)\right] + h(p\eta_1)\, h(p\eta_2) \,\kappa_e(p\eta)^{\phantom{\frac12}} \right\} + c.c., \nonumber \\  \quad \eta = \sqrt{\eta_1\, \eta_2},
\eqa
where the subscript $e$ of $n$, $\kappa$ and $\kappa^*$ stands to designate the exact (resumed) contribution.
Unlike the case of the principal series, this ansatz solves (\ref{DSDK2}) only in the region when all physical momenta in the above expressions are less than $\nu$. Hence, we substitute the asymptotic approximate expressions for all mode functions for the small values of their arguments.
In fact, this region brings the leading IR contributions to the integrals in (\ref{DSDK2}).

Thus, keeping only the leading contributions, we obtain from (\ref{ansatz0}) the following expression for the Keldysh propagator:

\bqa\label{DKIR}
D^{K}\left(p\, |\eta_1,\eta_2\right) \approx  A_-^2 \, \eta^{D-1} \, \frac{N(p\eta)}{\left(p\eta\right)^{2\nu}},
\eqa
where $N(p\eta) = 1 + 2 n_e(p\eta) + \kappa_e(p\eta) + \kappa_e^*(p\eta)$ in terms of the original $n_e$, $\kappa_e$ and $\kappa_e^*$. We assume that the initial condition for this quantity does not contain $\kappa$ and $\kappa^*$, but it can contain $n$. As we have explained in the Introduction, we put this condition at some initial value of the physical momentum, $n_0(p\eta) = n\left(P_0\right)$. This is the initial density perturbation on top of the BD state.

There are two points which are worth stressing at this moment. First, for generic values of $N(p\eta)$ the propagator $D^K\left(p\, |\eta_1,\eta_2\right)$ is not a function of the geodesic distance. That is true although combination $p\eta$ respects a part of the dS--isometry --- e.g. the simultaneous rescalings $p \to \sigma \, p$ and $\eta \to \eta/\sigma$.  Thus, any initial value for $N(p\eta)$ different from 1, violates the dS isometry. Second, substituting at the RHS of \eqref{DSDK2} the  tree-level value  $N(p\eta)$ =  1 would reproduce the  two--loop contribution to $D^K$ obtained above in the case when all modes are approximated by their values at $q_{1,2}\, \eta_{3,4} \ll \nu$.

Now we can substitute expressions \eqref{DRAIR} and \eqref{DKIR} at the RHS of \eqref{DSDK2}. The leading contributions will be given by the first two and last two terms of (\ref{DSDK2}). The other terms give rise to expressions which are suppressed by higher powers of $p\eta \to 0$. Also, it is convenient to make a change of variables $u = p\sqrt{\eta_1\eta_2}$, $v = \sqrt{\frac{\eta_3}{\eta_4}}$, and $\vec{l}_i = \vec{p}_i \sqrt{\eta_3\eta_4}$. Finally, we get

\bqa\label{DSDK21}
N(p\eta) \approx N\left(P_0\right) - \frac{\lambda^2}{3}A_-^6 A_+^2\int\limits^{p\eta}_{\nu}\frac{du}{u} \, \left[N(u)+N\left(P_0\right)\right] \, \int\limits^\nu_{\frac1\nu} \frac{dv}{v}\int\limits^{p\eta <\left|\vec{l}_{1,2}\right|< \nu} \frac{d^{D-1}\vec{l_1}}{(2\pi)^{D-1}}\frac{d^{D-1}\vec{l_2}}{(2\pi)^{D-1}} \times \nonumber \\ \times
\left[ \, \theta(v-1) \frac{1}{v^{2\nu}}- \theta(1-v) v^{2\nu}\right] \, \Biggl\{ \Biggr.
3 A_-^2 \,\frac{N(l_1)}{l_1^{2\nu}} \frac{N(l_2)}{l_2^{2\nu}}\left[v^{2\nu}-\frac{1}{v^{2\nu}}\right]
- A_+^2\,\left[v^{2\nu}-\frac{1}{v^{2\nu}}\right]^3\Biggr\},
\eqa
where $N\left(P_0\right)$ is the initial value of $N(p\eta)$ and $P_0 \sim \nu \gg p\eta$. In (\ref{DSDK21}) we neglected $p$ in comparison with $q_i$; this is made possible by the condition $D-1-4\nu > 0$.
This equation can be recast in the following form:
\bqa\label{main}
N(p\eta) - N\left(P_0\right) \approx - \int\limits^{p\eta}_\nu \frac{du}{u} \,\left[N(u) + N\left(P_0\right)\right] \,\left[ \Gamma_1 \left(\int\limits_{p\eta}^\nu dl \, l^{D-2 - 2\nu}\, N(l)\right)^2 - \Gamma_2\right],
\eqa
where
\bqa
\Gamma_1 = \frac{ \lambda^2 A_-^8 A_+^2 \, S^2_{D-2}}{\left(2\pi\right)^{2(D-1)}}  \,  \int\limits^\nu_{\frac1\nu} \frac{dv}{v}\left[ \, \theta(v-1) \frac{1}{v^{2\nu}}- \theta(1-v) v^{2\nu}\right]
\left[v^{2\nu}-\frac{1}{v^{2\nu}}\right] >0,
\notag\\
\quad{\rm and}\quad\Gamma_2 = \frac{\lambda^2 A_-^6 A_+^4 \, S^2_{D-2}}{3\, \left(2\pi\right)^{2(D-1)}}  \, \int\limits^\nu_{\frac1\nu}  \frac{dv}{v} \left[ \, \theta(v-1) v^{-2\nu}- \theta(1-v) v^{2\nu}\right] 	\,\left[v^{2\nu}-v^{-2\nu}\right]^3 >0.
\eqa
Here $S_{D-2}$ is the area of the $(D-2)$--dimensional sphere\footnote{First, the same equation is obtained for $\kappa^*(p\eta)$ instead of $N(p\eta)$, by using the out--modes instead of the BD modes  in (\ref{DSDK2}).}.  The above  equation can be transformed into an integro--differential equation by differentiating both sides w.r.t. $\log (p\eta)$:
\bqa
\frac{\partial N(p\eta)}{\partial \log\left(\frac{p\eta}{\nu}\right)} \approx - \left[N(p\eta) + N\left(P_0\right)\right] \, \left[\Gamma_1 \, \left(\int\limits^\nu_{p\eta} dl \, l^{D-2-2\nu} \, N(l)\right)^2 - \Gamma_2\right] + \nonumber \\ + 2\, \Gamma_1 \, N(p\eta) \, \left(p\eta\right)^{D-1-2\nu} \, \int\limits_\nu^{p\eta} \frac{du}{u} \, \left[N(u) + N\left(P_0\right)\right] \, \int\limits^\nu_{p\eta} dl l^{D-2-2\nu} \, N(l).
\eqa
Unlike the case of principal series \cite{Akhmedov:2013vka}, this equation has no  clear kinetic/particle interpretation. That is because in this case the modes do not oscillate at future infinity. Hence, we cannot neglect the time dependence of $N$ in comparison with that of the modes: here $N(p\eta)$ is just not a slow function. However, solving the equation for given initial conditions provides the resummation of the leading corrections from all loops (for the considered initial conditions).

In the limit  $p\eta \ll \nu$ let us consider
\bqa\label{smoothsol}
N(p\eta) = C (p\eta)^\alpha
\eqa
where $C$ is a constant of integration which depends on the initial conditions. Since $p\eta \ll P_0 \sim \nu$ for $\alpha > 0$ we have $N(p\eta) \ll N\left(P_0\right)$ and it can be easily seen that (\ref{smoothsol}) cannot solve (\ref{main}). But when  $\alpha < 0$ then $N(p\eta) \gg N\left(P_0\right)$. Hence, substituting into (\ref{main}) and neglecting $N\left(P_0\right)$ in comparison with $N$, we obtain that:
\bqa\label{Calpha}
\alpha \approx - \Gamma_1 \,  C^2 \, \left[\frac{\nu^{\left(D - 1 - 2\,\nu + \alpha\right)} - \left(p\eta\right)^{(D-1-2\nu + \alpha)}}{D - 1 - 2\, \nu + \alpha}\right]^2 + \Gamma_2.
\eqa
If $D-1-2\nu + \alpha > 0$ and $p\eta \ll \nu$, on the RHS of this equation we can neglect the $p\eta$ dependence and $\alpha$ is a constant, as it should be. This solution is valid only for
\bqa
 \, \frac{C \nu^{\left(D - 1 - 2\,\nu + \alpha\right)}}{D - 1 - 2\, \nu + \alpha} > \sqrt{\frac{\Gamma_2}{\Gamma_1}} ,
\eqa
(because otherwise $\alpha > 0$).

For the solution under consideration we have that $D^K(p|\eta_1, \eta_2) \approx \frac{A_-^2 \, C^2}{p^{D-1}} \, (p\eta)^{D-1-2\nu + \alpha}$. The Keldysh propagator blows up only if $D-1-2\nu+\alpha < 0$, but this cannot happen for the solution in question. Thus, all solutions of the type under consideration describe smooth behavior of the correlation functions, even if they violate dS isometry. Such solutions are realized by a mild initial perturbation over the BD state. Note that such a solution is very similar to the one obtained in \cite{Gautier:2015pca}--\cite{Serreau:2013psa}, if one keeps in the latter only the leading term as $p\eta \to 0$ (and we do keep only the leading contributions in the limit in question).
Furthermore, we have here an obvious stationary solution, $\alpha = 0$, $N(p\eta) = N\left(P_0\right) = C = \sqrt{\frac{\Gamma_2}{\Gamma_1}} \, \frac{D-1-2\nu}{\nu^{D-1-2\nu}}$.

It is tempting to compare the solution (\ref{smoothsol}), (\ref{Calpha}) to the one considered in \cite{Gautier:2015pca}, \cite{Serreau:2013psa}. There are certain differences. Namely our solution is valid for generic dS violating initial conditions. As the result, the parameter $\alpha$ in (\ref{smoothsol}) depends on the parameter $C$, which defines initial value of $N(p\eta)$. The situation should to reduce to the one considered in \cite{Gautier:2015pca}, \cite{Serreau:2013psa}, when one takes exactly BD state. In terms of (\ref{smoothsol}) that corresponds to $N(\nu) = 1$. In this case, as follows from (\ref{Calpha}), $\alpha \sim \frac{\lambda^2}{m^4}$, if $m^2 \ll 1$ (this is the approximation, in which one can compare the two results under discussion). This answer coincides parameterically with the one found in \cite{Gautier:2015pca}, \cite{Serreau:2013psa}. The coefficients, however, are different. The reason for that is due to the difference between the sort of approximations made and sort of leading diagrams that are resummed in \cite{Gautier:2015pca}, \cite{Serreau:2013psa} and in our paper.

However, apart from the stable solutions eq. (\ref{main}) also has a singular (exploding) one. Consider indeed
\bqa\label{blowup}
N(p\eta) = \frac{C}{\left(p\eta - p\eta_*\right)^\alpha},
\eqa
where $C$, $0 < \eta_* < \eta$ and $\alpha > 0$ are some real constants, which may depend on the initial conditions. We assume that such a behavior of $N(p\eta)$ is valid in the limit when $\eta$ is very close to $\eta_*$. After the substitution of this solution into (\ref{main}) and neglecting the suppressed terms, we obtain the following relation between the constants $C$, $\eta_*$ and $\alpha$:

\bqa
\frac{1}{\left(p\eta - p\eta_*\right)^\alpha} \approx \frac{\Gamma_1 \, C^2 \, \left(p\eta_*\right)^{D-1-2\nu}}{(\alpha - 1) \, \left(p\eta - p\eta_*\right)^{3(\alpha - 1)}}.
\eqa
This equation establishes that $\alpha = 3/2$ and a relation between $C$ and $\eta_*$. In this case the Keldysh propagator blows up at a finite proper time\footnote{It is probably worth stressing at this point that here we are talking about superhorizon modes and, hence, even for the very short periods of times the intuition gained from the flat space-time physics is not applicable here.}. Then, also  the expectation value of the stress--energy tensor blows up (which would appear at the RHS of the Einstein equations due to the quantum fluctuations). That means that the backreaction is not negligible. One possibility is that  that cosmological constant is secularly screened because the expectation value of the stress--energy tensor under discussion does not respect the dS isometry. This is a subject of a separate study. Here we do not consider the backreaction issue.

\section{Conclusions and acknowledgements}

Eq. (\ref{DKBD}) shows that there are secularly growing corrections to the Keldysh propagator starting from two loops order --- i. e.  from the sunset diagrams. As explained in  Section \ref{IIIA} this equation is valid for $\frac{\sqrt{5}}{6}\, (D-1) < m < \frac{D-1}{2}$.  Moreover, as  explained in Section \ref{IIIB}, for  $m < \frac{\sqrt{3}}{4}\, (D-1)$ there are large IR contributions to the vertices. The physical origin of these complications  is explained n the section \ref{IV}.

Since the quantum corrections to the propagator become of the same order as tree--level contributions, when $\left|\lambda^2 \, \log(p\eta)\right| \sim 1$ they have to be resummed. We perform a self--consistent resummation, by resumming only the leading corrections in $\lambda^2 \log(p\eta)$ and dropping  all the subleading ones --- those which are suppressed by higher powers of $\lambda$ or do not contain $\log(p\eta)$, as $p\eta \to 0$.

The resummation amounts to solving the relevant Dyson--Schwinger equation. For the Keldysh propagator we make the ansatz (\ref{DKIR}) with unknown $N(p\eta)$, and we take tree--level (perhaps UV renormalized) expressions for the retarded progators, the advanced propagators and the vertices because they do not receive large IR contributions at leading order for the mass range under consideration.

Eq. (\ref{main}) follows from the Dyson--Schwinger equation.
Its solution allows to perform the resummation of leading secular IR corrections from all loops. Note that  $N(p\eta)$ is a quantity attributed to the comoving volume; its physical meaning is however less clear than in the principal series case. However, we can solve (\ref{main}) and find the behavior of the Keldysh propagator at future infinity.

The solution (\ref{smoothsol}) and (\ref{Calpha}) describes the smooth behavior of the Keldysh propagator and corresponds just to a mass renormalization. This situation is very similar to the one encountered in \cite{Gautier:2015pca}--\cite{Serreau:2013psa}.

However, the selfconsistent resummation  (\ref{main}) produces a  nonlinear equation. As a result it also has the exploding solution (\ref{blowup}). Which one of the solutions is realized depends on the initial conditions $N(P_0)$. The blow up happens at finite proper time and cannot be washed away by the EPP expansion, because $N(p\eta)$ is attributed to the comoving volume.

In conclusion, in dS space the backreaction on quantum effects can be strong also for massive fields (see also \cite{Akhmedov:2013vka}).
\vskip20pt

We would like to acknowledge discussions with H. Godazgar, A.Polyakov, J. Serreau and V. Slepukhin. ETA would like to thank AEI, Golm for hospitality during the final stage of work on this project. The work of ETA and FKP was partially supported by the grant for the support of the leading scientific schools SSch--1500.2014.2, by their grants from the Dynasty foundation and was done under the partial support of the RFBR grant 15-01-99504.
FKP is grateful to the grant of RFBR 16-32-00064-mol-a and the Ministry of
Education and Science of the Russian Federation (Contract No.
02.A03.21.0003 dated of August 28, 2013) for travel support. The work of ETA was supported by the state grant Goszadanie 3.9904.2017/BCh.

\end{document}

$$\phi(\eta, \vec{x}) = \eta^{(D-1)/2}\,h(p\eta)\, e^{-i\,\vec{p}\, \vec{x}} a(p) + \eta^{(D-1)/2}\,h^*(p\eta)\, e^{i\,\vec{p}\, \vec{x}} a^+(p) $$